# Electrophilicity as a Possible Descriptor of the Kinetic Behaviour

**P. K. Chattaraj*** and **U. Sarkar**

*Chemistry Department, Indian Institute of Technology*

*Kharagpur-721302, India*

and

**M. Elango**, **R. Parthasarathi** and **V. Subramanian***

*Chemical Laboratory, Central Leather Research Institute,*

*Adyar, Chennai- 600 020, India.*

**ABSTRACT**: An understanding of the precise nature of a chemical reactivity descriptor is of utmost interest to quantum chemists. In the present investigation an attempt has been made to analyze whether the electrophilicity index is a reliable descriptor of the kinetic behaviour. Relative experimental rates of Friedel- Crafts benzylation, acetylation and benzoylation reactions correlate well with the corresponding calculated electrophilicity values. Chlorination of various substituted ethylenes and nitration of toluene and chlorobenzene are studied as representative examples of electrophilic addition and substitution reactions respectively. The correlation is not very good to assess that which, however, improves drastically by removing a few data points to show that the electrophilicity is a kinetic quantity with inherent thermodynamic information. The correlation between the experimental and the calculated activation energies is studied for some Markovnikov and anti-Markovnikov addition reactions and is turned out to be reasonably well. Reaction electrophilicity, local electrophilicity and activation hardness are used together to provide a transparent picture of reaction rates as well as the orientation of aromatic electrophilic substitution reactions. Ambiguity in the definition of the electrophilicity is highlighted.

*Authors for correspondence: E-mail: pkc@chem.iitkgp.ernet.in, subuchem@hotmail.com

## Introduction

Many of the organic reactions can be described in terms of the electro (nucleo) philic addition and substitution. These reactions have got large synthetic potentials and are most widely studied [1, 4]. Traditionally the electrophilicity is treated [1, 4] as a kinetic quantity, which explains the rate of a reaction through its correlation with activation energy or free energy of activation occasionally supplemented by thermodynamic stabilities of various species involved. On the other hand, the nucleophilicity and the basicity are often analyzed at par [1, 4] since both involves the amount of electron density present in it and its potential to donate that, although the former correlates with $\Delta G^{\ddagger}$ (a kinetic quantity) and the latter with $\Delta G_r$ (an equilibrium or thermodynamic property). Although it has been known for a long time that the electrophilicity is a cardinal index of reactivity and selectivity, an acceptable definition of it was lacking. Based on the work of Maynard et al [5] a theoretical definition of electrophilicity has been introduced recently by Parr et al. [6] It may be noted that Maynard et al [5] and Parr et al [6] have prescribed the same definition of electrophilicity through essentially kinetic (via correlation with reaction rates) and thermodynamic (in terms of the energetically favorable charge transfer processes) routes respectively and hence it is expected that it will contain both kinetic and thermodynamic information. This electrophilicity, however does not correlate well with the electron affinity [6].

Density functional theory [7, 8] has been quite successful in providing theoretical background of popular qualitative chemical concepts. In this context, several reactivity descriptors have been proposed and used to analyze chemical reactivity and site selectivity. Hardness, global softness, electronegativity and polarizability are the global reactivity descriptors widely used to understand the global nature of molecules in terms of their stability and it is possible to gain knowledge about the reactivity of molecules. Atomic charges, Fukui functions (FF) and local softness are the local reactivity descriptors, which provide information about the site selectivity. In addition to these reactivity descriptors, Hard and Soft Acids and Bases (HSAB) principle has been employed in number of cases in analyzing both nucleophilicity and basicity, which encapsulates both thermodynamic and kinetic properties of numerous molecules [9, 10]. A hard (soft) nucleophile prefers to react with a hard (soft) electrophile for both kinetic and thermodynamic considerations and for two species of comparable electronegativity values [9, 10].

Zhou and Parr [11] have defined the activation hardness and investigated the electrophilic substitution of aromatic compounds using that. It is expected that the electro (nucleo) philicity should have both kinetic and thermodynamic requirements. The main objective of the present work is to gain insights into the exact nature (kinetic or thermodynamic) of the electrophilicity index. Different types of Friedel- Crafts reactions like benzylation, acetylation and benzoylation are studied to correlate the experimental rates of those reactions with the corresponding theoretical electrophilicity values. Reliability of calculated activation energies and the problems associated with the definition of electrophilicity for more than one species are also discussed. Chlorination of various substituted ethylenes and nitration of toluene and chlorobenzene are taken as the representative reactions encompassing the electrophilic addition and substitution respectively.

## Theoretical Background

Various global and local reactivity descriptors used in the present work are described below.

*Activation hardness:* Zhou and Parr [11] have proposed the activation hardness in accordance with the transition state (TS) theory. The activation hardness is defined as

$$\eta^{\ddagger} = \eta_{R} - \eta_{TS} \tag{1}$$

where $\eta^{\ddagger}$ is the activation hardness, $\eta_R$ is the hardness of the reactant and $\eta_{TS}$ is the hardness of the transition state. Zhou and Parr [11] have shown that smaller the activation hardness, the faster is the reaction.

One of the difficulties associated with the calculation of hardness of the reactants is that the hardness of the chemical species is not additive. The hardness of two chemical species, *viz.,* A, the acceptor and D, the donor is calculated by the relationship [11]

$$\eta = \frac{(I_D - A_A)}{2} \tag{2}$$

where $I_D$ is the ionization potential of the donor and $A_A$ is the electron affinity of the acceptor.

*Electrophilicity:* Electrophilicity index ($\omega$) has been defined by Parr et al [6] as

$$\omega = \frac{\mu^2}{2\eta} \tag{3}$$

In Eq. (3) $\mu \approx$ -(I+A)/2 and $\eta \approx$ (I-A)/2 are the electronic chemical potential and the chemical hardness respectively, approximated in terms of the vertical ionization potential (I) and electron affinity (A). The electrophilicity is a descriptor of reactivity that allows a quantitative classification of the global electrophilic nature of a molecule within a relative scale and effectively is the power of a system to ‘soak up’ electrons [6].

The local version of the electrophilicity index has been proposed by employing a resolution of identity as follows [12]:

$$\omega_k^\alpha = \omega f_k^\alpha \tag{4}$$

where $f_k$ is the Fukui function at atom k in a molecule and ($\alpha$= +, - and 0) represents local philic quantities describing nucleophilic, electrophilic and radical attacks respectively. Since, electrophilicity measures the energy change of an electrophile as it is saturated with electrons, it may be considered to be an additive parameter. This property of electrophilicity has been used in this study to define new reactivity quantities such as activation electrophilicity and reaction electrophilicity.

*Activation electrophilicity:* Consider the following chemical reaction

$$A + B \rightarrow C + D \tag{5}$$

Let $\omega_A$ ,$\omega_B$, $\omega_C$ and $\omega_D$ be the global electrophilicity indexes of reactants A, B and products C, D respectively. Considering this reaction to proceed via a transition state, it is possible to define activation electrophilicity by

$$\omega^{\ddagger} = \omega_{TS} - \omega_R \tag{6}$$

where $\omega_R = \omega_A + \omega_B$.

The corresponding activation energy is given by

$$E^{\ddagger} = E_{TS} - E_R \tag{7}$$

where $E_R = E_A + E_B$

$\omega_R$ and $E_R$ are the electrophilicity index and energy of reactants respectively, and $\omega_{TS}$ and $E_{TS}$ are the electrophilicity index and energy of the transition state respectively.

*Reaction electrophilicity:* Reaction electrophilicity is defined as

$$\omega_r = \omega_P - \omega_R \quad (8)$$

where $\omega_P = \omega_C + \omega_D$.

The corresponding reaction energy is given as

$$E_r = E_P - E_R \quad (9)$$

where $E_P = E_C + E_D$ and

$\omega_P$ and $E_P$ are the electrophilicity index and energy of products respectively.

For the reactions involving more than one reactant or product the definition of electrophilicity becomes ambiguous. We consider here some of the probable definitions. For a reaction of the type: A+B $\rightarrow$ TS $\rightarrow$ C+D we may define the electrophilicity of the reactant as:

$$\omega_R^{(1)} = \omega_A + \omega_B \quad (10a)$$

$$\omega_R^{(2)} = \omega_A, \text{ if B is common for a series of A type molecules} \quad (10b)$$

$$\omega_R^{(3)} = \frac{\mu_{AB}^2}{2\eta_{AB}} = \frac{\chi_{AB}^2}{2\eta_{AB}} \quad (10c)$$

where
$$\chi_{AB} = \frac{I_{\min} + A_{\max}}{2} \quad (10d)$$

and
$$\eta_{AB} = \frac{I_{\min} - A_{\max}}{2} \quad (10e)$$

Three different types of electrophilicity for the products can also be defined accordingly.

## Computational Details

General reaction scheme for all the reactions considered in the study is presented in scheme I. All the geometries of the molecules concerned are optimized in gas phase using Becke's three parameter hybrid density functional, [13] B3LYP/6-31G*, which includes both Hartree-Fock exchange and DFT exchange correlation functionals [14,15] using the Gaussian 98W [16] package. The minimum energy configurations of the reactants, intermediates and products and the saddle point nature of the transition states have been ensured with the help of the corresponding calculated vibrational frequencies. The reactants, intermediates and products are associated with zero imaginary frequencies, whereas there is one imaginary frequency for each transition state. Reactivity descriptors like chemical hardness and electrophilicity index have been calculated using standard working equations described earlier. AIM analyses for the π-complexes are carried out with the help of AIM 2000 software package [17]. The thermodynamic parameters for various reactions have been computed by the standard method implemented in the Gaussian package [16]. Using the freq keyword, the free energies of various reactions have been computed at T= 298.15K. From the calculated free energies of the transition states ($G_{TS}$) and those of reactants ($G_R$), the free energy of activation have been obtained by $\Delta G^{\ddagger} = G_{TS} - G_R$. From the free energies of products ($G_P$) and the reactants ($G_R$), the free energy of reaction have been calculated using the following equation $\Delta G_r = G_P - G_R$.

## Results and Discussion

*Friedel- Crafts reaction:* Optimized structures of all the molecules involved in the benzylation, acetylation and benzoylation reactions studied here are collected in Figure 1, but not shown due to space limitations. Experimental relative rates $\left(RR = \frac{k_{toluene}}{k_{benzene}}\right)$ are taken from reference [18] and are correlated with the calculated electrophilicity values. Table I presents the theoretical ω values along with the experimental and calculated $\ln(RR)$ values.

As we see in Figure 2, the experimental $\ln(RR)$ values for none of the benzylation, acetylation and benzoylation reactions correlate very well with the ω values. However, the correlation improves significantly (R=0.928, 0.898, 0.993) once we neglect some of the points. Mainly the nitro- and chloro- substituted compounds in the benzylation reactions and highly fluoro- substituted and crowded molecules in the benzoylation reactions exhibit different behaviour than the rest. It may, however, be noted that there is no a priori method known to judge those points causing the "discrepancies" and to provide a rationale for that. Perhaps it stems from the quality of the experiments and/or the definition of electrophilicity (to especially take care of the kinetic aspects) and its method of calculation. Currently we are exploring the dependence of the reaction rates on both the global and the local electrophilicities. Preliminary results are very encouraging. This may allow us to consider the electrophilicity as a kinetic quantity and may be used in estimating the rate of a chemical reaction and consequently the associated Hammett constant and the nucleus independent chemical shift, wherever applicable [19].

It is not always easy to gather the experimental rates and reaction energies of several reactions to come to a definite conclusion. In order to bypass this problem we propose to calculate the activation and reaction energies of some reactions. For gaining confidence we first calculate the activation energies of some Markovnikov and anti-Markovnikov addition reactions of hydrogen halides to alkenes whose experimental activation energies are known [20]. Table II presents the activation and reaction energies as well as the experimental activation energies obtained from a synchrotron radiation experiment [20]. A linear correlation (Figure 3, R=0.970) between the experimental and the theoretical activation energies provides confidence and helps us to proceed further with the corresponding theoretical quantities for the following reactions.

*Chlorination of various substituted ethylenes:* In chlorination of various substituted ethylenes, chlorine acts as an electrophile and π-electrons of the substituted ethylene are the nucleophile and the gas phase reaction proceeds via a cyclic chloronium bridged ion [1, 4]. The substituted ethylene and chlorine form a π-complex before forming the transition state. The π-electrons attack the chlorine and displace a chlorine ion to form a cationic cyclic chloronium ion as an intermediate. This step proceeds via a transition state that involves breaking of the Cl- Cl bond as well as breaking the π-bond in the substituted ethylene in a concerted manner. Formation of chloronium ion is the slow and the rate-determining step. Therefore, the transition state formed in the rate-determining step is of much interest in calculating activation quantities.

Figure 4 is reserved for the structures and selected bond lengths for these transition states (not shown because of space constraints). Chlorine ($Cl_2$) and substituted ethylenes ($CH_2CHR$) are the reactants. Final anti product is the product considered for calculating reaction quantities. Various substituents on ethylene influence the rate of electrophilic addition to it. It is well known that electron-withdrawing substituents decrease the reactivity whereas electron-donating ones increase the reactivity.

In the present study the existence of a weak complex between substituted ethylenes and $Cl_2$ have been analyzed with the help of an atoms-in-molecule (AIM) approach developed by Bader and his group [21] by locating the critical points at which the gradient of electron density vanishes. The topographical features, *i.e.* the presence of bond critical points between the C atom and $Cl_2$, the value of the electron density at the bond critical points ($\rho(\vec{r}_c)$) and the Laplacian of the electron density at the bond critical points ($\nabla^2 \rho(\vec{r}_c)$) have revealed the existence of weak complexes before the formation of the transition state. The calculated $\rho(\vec{r}_c)$, $\nabla^2 \rho(\vec{r}_c)$ and the interaction energy ($E_{Int}$) for various $\pi$ complexes are shown in Table III.

The calculated $\rho(\vec{r}_c)$ and $\nabla^2 \rho(\vec{r}_c)$ values are correlated with the interaction energy calculated using the following formula,

$$E_{Int}=E_{cpx}- (E_{alkene} +E_{halogen})$$

Interaction energy correlated well with $\rho(\vec{r}_c)$ and $\nabla^2 \rho(\vec{r}_c)$ with R values of 0.996 and 0.973 respectively, and the respective plots are presented in Figure 5. Among all the substituted ethylenes studied, the $NO_2$- and NO- substituents are the most deactivating systems towards an electrophilic addition reaction. For these two cases the $\pi$- complex formation with $Cl_2$ was not observed.

In order to verify whether electrophilicity is a kinetic or a thermodynamic quantity the activation energy ($E^{\ddagger}$), reaction energy ($E_r$), activation free energy ($\Delta G^{\ddagger}$) and free energy of reaction ($\Delta G_r$) are calculated and reported in Table IV. Electrophilicity ($\omega$) values of the reactant, the transition state and the product are presented in Table V. Their correlations with various kinetic and thermodynamic quantities (reported in Table IV) are provided in Figure 6. It is observed that the electrophilicity does not correlate well with either the thermodynamic or the kinetic quantities. However, the correlation improves substantially by neglecting some odd points.

The inter correlation patterns between the kinetic and thermodynamic sets are shown in Figure 7 wherein it becomes transparent that they themselves correlate to some extent among each other.

Using equations (1), (6) and (8) activation hardness ($\eta^{\ddagger}$), activation electrophilicity ($\omega^{\ddagger}$) and reaction electrophilicity ($\omega_r$) are calculated and they are presented in Table VI. $E^{\ddagger}$ is a predictor of relative rates and it is a kinetic quantity and any other quantity that correlates well with $E^{\ddagger}$ and $\Delta G^{\ddagger}$ can be considered as a kinetic quantity. Hence, an attempt has been made to correlate $\omega_r$ and $\omega^{\ddagger}$ with $E^{\ddagger}$ and $\Delta G^{\ddagger}$. The possible relationships between various kinetic quantities with $\omega_r$ are shown in Figure 8. In order to derive information about the nature of $\eta^{\ddagger}$, the possible linear regression with $E^{\ddagger}$ and $\Delta G^{\ddagger}$ is also attempted and are depicted in Figure 9.

It is evident from Figure 8 that reaction electrophilicity exhibits a poor linear relationship with $E_r$, $E^{\ddagger}$, $\Delta G_r$ and $\Delta G^{\ddagger}$. A drastic improvement is noticed (on the right hand side of each plot) by omitting a few points. It is noticeable that $\omega_r$ correlates well with $E^{\ddagger}$ and $\Delta G^{\ddagger}$ and thereby confirming that $\omega_r$ is essentially a kinetic parameter. However, $\omega^{\ddagger}$ does not show any linear relationship with these kinetic quantities. It may be noted that the $R$ values for the plots of $E_r$ and $\Delta G_r$ versus $\omega_r$ are not negligible and almost comparable to that of the plots of $E^{\ddagger}$ and $\Delta G^{\ddagger}$ versus $\omega_r$ which confirms the additional thermodynamic information content of $\omega_r$. Similarly, $\eta^\ddagger$ shows good linear relationship with $E^\ddagger$ and $\Delta G^\ddagger$ (Figure 9). The $R$ values suggest that the activation hardness is also a kinetic parameter.

*Nitration of toluene and chlorobenzene:* In an electrophilic aromatic substitution reaction the electrophile replaces a proton from the ortho, meta or para position or another Lewis acid leaving group from the ipso position. The electrophile in the nitration reactions is the nitronium ion ($NO_2^+$) formed via reaction of $HNO_3$ and $H_2SO_4$. The reaction proceeds [1, 4] through any of the three possible intermediates, *viz*. two types of $\pi$ - complexes and a $\sigma$ - complex, called a benzenium ion. All of these intermediates are stable species as vindicated by their *nmr* spectra. It has been long believed that the transition states (late) associated with electrophilic aromatic substitutions resemble the $\sigma$ - complexes more than the two types of $\pi$ - complexes and the formation of the $\sigma$ - complex is the rate determining step in most cases and is irreversible for all practical purposes in many reactions. This leads to the general rules governing these reactions as [1] a) Electron donating (-I) groups increase the rates (activate by stabilizing the TS for the $\sigma$ - complex formation) and direct the electrophile predominantly to the ortho- or the para- positions and b) Electron withdrawing (+I) groups decrease the rates (deactivate by destabilizing the TS for the $\sigma$ - complex formation) and direct the electrophiles predominantly to the meta- positions. Of course the resonance effects ($\pm$M) of the substituents are also to be considered in understanding their effects on rates. It may be noted that in other text books (e.g. references [3] and [4]) electron donating and withdrawing effects are designated as (+I, +M) and (-I, -M) respectively. At high temperature all groups show preferences towards the meta- positions mainly because of thermodynamic control. It is expected that the reactions are effectively kinetically controlled at the temperature (298.15 K) in which the free energies are calculated. Partial rate factors are argued [1, 4] to be better descriptors of the substituent effects on rates than their relative rates. Both $–CH_3$ and –Cl groups are ortho- para directing although the former is activating (-I) and the latter is deactivating (+I) but having pronounced resonance effects. Various transition states and $\sigma$ - complexes associated with the nitration of toluene and chlorobenzene are stored in Figure 10 (not shown). Table VII reports the electrophilicity values of all the species involved in these reactions. Their relative energies and electrophilicities are provided respectively in Figures 11 and 12. It is important to note that the transition states and the intermediates are always more electrophilic than the reactants and the products.

Table VIII presents $\omega_r$, $E_r$, $\omega^{\ddagger}$, $E^{\ddagger}$ and $\eta^{\ddagger}$ of these species. It is clear from the $E^{\ddagger}$ values that both $–CH_3$ and –Cl will be o-p directing. Although the $\eta^{\ddagger}$ values properly take care of $–CH_3$ it fails in case of –Cl. In case of intramolecular reactivity, $\omega_r$ does not correlate well with $E^{\ddagger}$ which may be due to the fact that the reactant is the same in all cases. However, $\omega_r$ (like $E^{\ddagger}$) values of the m-products are larger than those of the o-p products in both cases.

The philicity ($\omega_k^-$) values of toluene and chlorobenzene are reported in Table IX. They clearly reveal that both $–CH_3$ and –Cl are o-p directing and are found to be better descriptors of orientation of aromatic electrophilic substitutions than $\eta^{\ddagger}$. Note that the respective Fukui functions will suffice in case the intramolecular reactivity is considered.

## Concluding Remarks

In order to test the potential of the electrophilicity as a descriptor of the kinetic characteristics, Friedel- Crafts benzylation, acetylation and benzoylation reactions, electrophilic addition to various substituted ethylenes and electrophilic aromatic substitution reactions in toluene and chlorobenzene are studied. It has been observed that if a few systems are neglected the electrophilicity correlates very well with the experimental rates and hence is essentially a kinetic concept but it has also inherited an adequate amount of thermodynamic information due to their intercorrelations. Different ways of defining electrophilicity for reactants and products are discussed. A linear correlation between the experimental and the calculated activation energies for some Markovnikov and anti-Markovnikov addition of hydrogen halides to alkenes are observed. Reaction electrophilicity complements activation hardness concept in understanding the rates of the electrophilic addition and substitution reactions and to some extent the stability of the associated products as well. Philicity also adequately describes the orientation of electrophilic aromatic substitution reactions.

## ACKNOWLEDGMENTS

We thank CSIR, New Delhi for financial support and Professors A. Basak and D. Mal for helpful discussions.

## References

1. Carey, F.; Sundberg, R. Advanced Organic Chemistry: Structure and Mechanism, 3rd Ed. Plenum Press: New York, 1993.
2. Lowry, T. H.; Richardson, K. S. Mechanism and Theory in Organic Chemistry, 3rd Ed. Harper Collins: New York, 1987.
3. March, J. Advanced Organic Chemistry: Reactions, Mechanisms and Structure, 4th Ed. John Wiley: New York, 1992.
4. Finar, I. L. Organic Chemistry: The Fundamental Principle, 6th Ed. English Language Book Society: London, 1990.
5. Maynard, A. T.; Huang, M.; Rice, W. G.; Covell, D. G. Proc Natl Acad Sci USA 1998, 95, 11578.
6. Parr, R. G.; Szentpaly, L. v.; Liu, S. J Am Chem Soc 1999, 121, 1922.
7. Parr, R. G.; Yang, W. Density Functional Theory of Atoms and Molecules, Oxford University Press: Oxford, 1989.
8. Geerlings, P.; De Proft, F.; Langenaeker, W. Chem Rev 2003, 103, 1793.
9. Pearson, R. G. Chemical Hardness - Applications from Molecules to Solids, VCH-Wiley: Weinheim, 1997.
10. Chattaraj, P. K.; Lee, H.; Parr, R. G. J Am Chem Soc 1991, 113, 1855.
11. Zhou, Z.; Parr, P. G. J Am Chem Soc 1990, 112, 5720.
12. Chattaraj, P. K.; Maiti, B.; Sarkar, U. J Phys Chem A 2003, 107, 4973.
13. Becke, A. D. J Chem Phys. 1993, 98, 5648.
14. Lee, C.; Yang, W.; Parr, R. G. Phys Rev B 1998, 37, 785.

15. Stephens, P. J.; Devlin, F. J.; Chabalowski, C. F.; Frisch, M. J. J Phys Chem 1994, 98, 11623.

16. Gaussian 98, Revision A.5, Frisch, M. J.; Trucks, G. W.; Schlegel, H. B. et al. Gaussian, Inc., Pittsburgh, PA, 1998.

17. F. Biegler-Konig, J. Schonbohm, R. Derdau, D. Bayles and R. W. F. Bader, AIM 2000, version 1; Bielefeld, Germany, 2000.

18. Olah, G. A. Acc Chem Res 1971, 4, 240.

19. Elango, M.; Parthasarathi, R.; Narayanan, G. K.; Sabeelullah, A. Md.; Sarkar, U.; Venkatasubramaniyan, N. S.; Subramanian, V.; Chattaraj, P. K. J Chem Sci 2005, 117, 1.

20. Sæthre, L. J.; Thomas, T. D.; Svensson, S. J Chem Soc Perkin Trans 1997, 2, 749.

21. Bader, R. F. W. Atoms in Molecules: A Quantum Theory, University of Oxford Press: Oxford, 1990.

## Benzylation

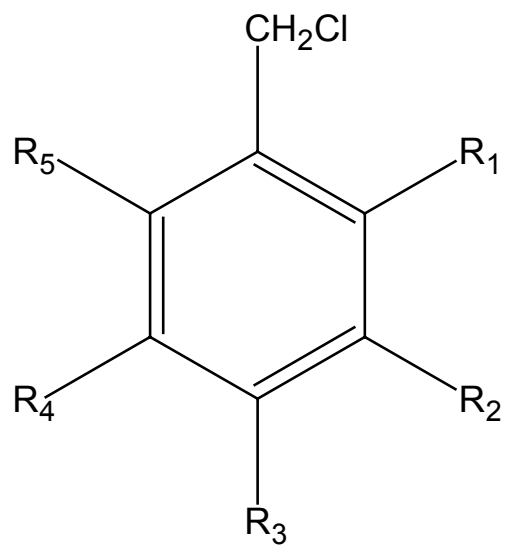

| | $R_1$ | $R_2$ | $R_3$ | $R_4$ | $R_5$ |
|---|---|---|---|---|---|
| 1 | H | H | $NO_2$ | H | H |
| 2 | F | H | H | H | H |
| 3 | H | F | H | H | H |
| 4 | H | H | F | H | H |
| 5 | Cl | H | H | H | H |
| 6 | H | Cl | H | H | H |
| 7 | H | H | Cl | H | H |
| 8 | H | H | H | H | H |
| 9 | $CH_3$ | H | H | H | H |
| 10 | H | $CH_3$ | H | H | H |
| 11 | H | H | $CH_3$ | H | H |
| 12 | $CH_3$ | H | $CH_3$ | H | $CH_3$ |
| 13 | $OCH_3$ | H | H | H | H |
| 14 | H | $OCH_3$ | H | H | H |
| 15 | H | H | $OCH_3$ | H | H |
| 16 | $OCH_3$ | H | $OCH_3$ | H | $OCH_3$ |

## Acetylation

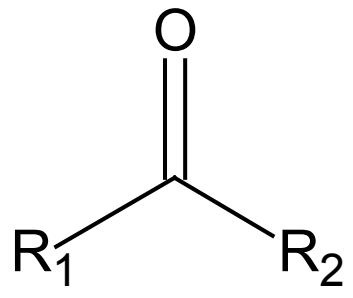

| | $R_1$ | $R_2$ |
|---|---|---|
| 1 | F | H |
| 2 | F | $OCH_3$ |
| 3 | Cl | $CH_3$ |
| 4 | Cl | $C_2H_5$ |
| 5 | Cl | $CH(CH_3)_2$ |
| 6 | Cl | $CH_2Cl$ |
| 7 | Cl | $CHCl_2$ |

## Benzoylatio

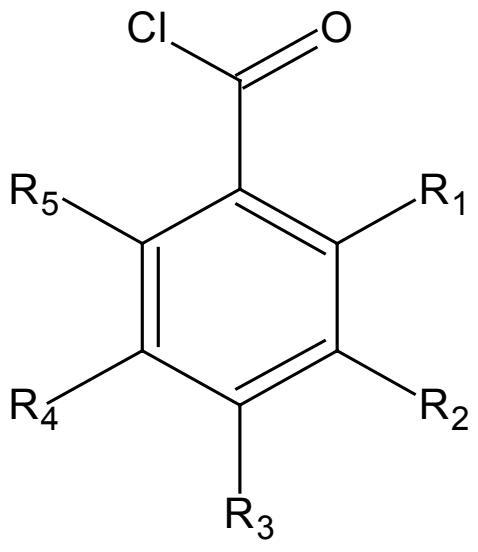

| | $R_1$ | $R_2$ | $R_3$ | $R_4$ | $R_5$ |
|---|---|---|---|---|---|
| 1 | F | F | F | F | F |
| 2 | $NO_2$ | H | $NO_2$ | H | H |
| 3 | H | $NO_2$ | H | $NO_2$ | H |
| 4 | H | H | $NO_2$ | H | H |
| 5 | F | H | H | F | H |
| 6 | H | H | H | H | H |
| 7 | H | H | $CH_3$ | H | H |
| 8 | H | H | F | H | H |
| 9 | $CH_3$ | H | $CH_3$ | H | $CH_3$ |
| 10 | H | H | $OCH_3$ | H | H |

## Addition of chlorine to alkenes

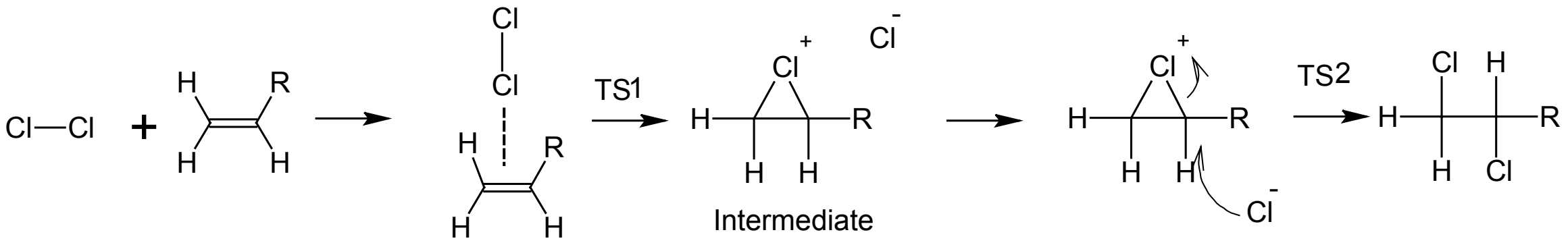

## Nitration of Toluene/chlorobenzene (only *ortho* nitration is shown)

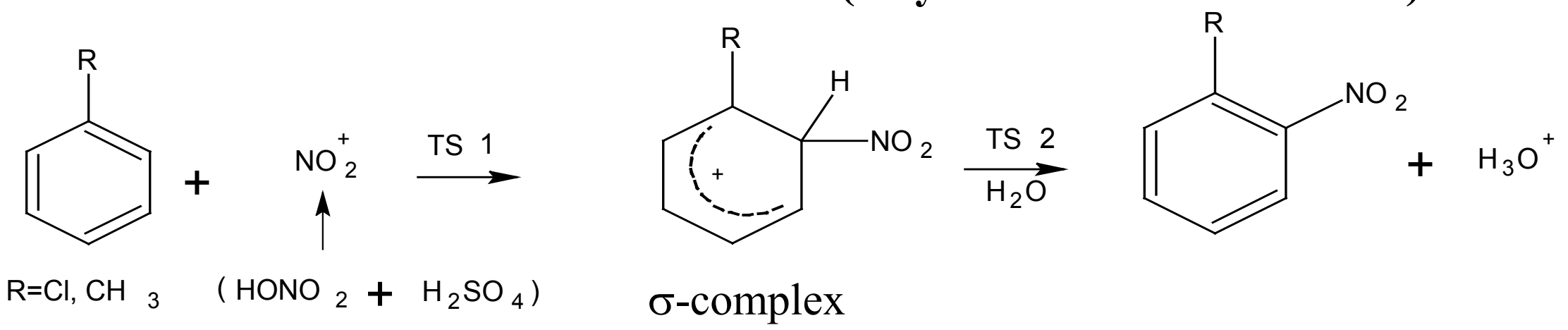

**Scheme 1** General reaction scheme for all the reactions chosen for the present investigation

**TABLE I**

Theoretical electrophilicity (ω), calculated ln(RR) and experimental ln(RR) values associated with the Friedel- Crafts benzylation, acetylation and benzoylation reactions.

| Molecules | ω (eV) | Calculated ln(RR) [a] | Experimental ln(RR) |
|---|---|---|---|
| Benzylation | | | |
| 1 | 5.369 | 0.213 | 0.916 |
| 2 | 2.581 | 2.511 | 1.569 |
| 3 | 2.628 | 2.472 | 1.526 |
| 4 | 2.412 | 2.650 | 2.163 |
| 5 | 2.180 | 2.841 | 1.526 |
| 6 | 2.753 | 2.369 | 1.856 |
| 7 | 2.726 | 2.391 | 1.825 |
| 8 | 2.396 | 2.664 | 1.841 |
| 9 | 1.742 | 3.203 | 2.950 |
| 10 | 2.307 | 2.737 | 2.054 |
| 11 | 2.250 | 2.783 | 3.367 |
| 12 | 2.198 | 2.826 | 3.666 |
| 13 | 2.086 | 2.918 | 4.099 |
| 14 | 2.111 | 2.898 | 2.580 |
| 15 | 1.945 | 3.035 | 4.575 |
| 16 | 2.091 | 2.915 | 4.913 |
| Acetylation | | | |
| 1 | 2.887 | 4.465 | 3.544 |
| 2 | 2.377 | 4.576 | 4.868 |
| 3 | 3.073 | 4.425 | 4.949 |
| 4 | 3.031 | 4.434 | 4.491 |
| 5 | 2.921 | 4.458 | 4.436 |
| 6 | 4.289 | 4.161 | 4.359 |
| 7 | 4.722 | 4.067 | 3.940 |
| Benzoylation | | | |
| 1 | 5.047 | 4.445 | 2.779 |
| 2 | 7.540 | 3.371 | 3.367 |
| 3 | 7.280 | 3.483 | 3.661 |
| 4 | 6.917 | 3.639 | 3.951 |
| 5 | 4.795 | 4.554 | 4.566 |
| 6 | 4.277 | 4.777 | 5.034 |
| 7 | 4.065 | 4.868 | 5.102 |
| 8 | 4.353 | 4.744 | 5.136 |
| 9 | 2.803 | 5.412 | 5.278 |
| 10 | 3.681 | 5.034 | 5.451 |

[a] Calculated ln(RR) = 4.63765-0.82405*ω for Benzylation
Calculated ln(RR) = 5.09156-0.21697*ω for Acetylation
Calculated ln(RR) = 6.61966-0.43089*ω for Benzoylation

**TABLE II**

Calculated activation and reaction energies and experimental activation energies of Markovnikov[m] and anti-Markovnikov[a] addition of hydrogen halides to alkenes. All the values are in kcal.mol$^{-1}$.

| Reactants | $E^{\ddagger}$ | $E_r$ | (Exp.) $E^{\ddagger}$ |
|---|---|---|---|
| Ethene:HF | 38.3 | -23.05 | 49.1 |
| Ethene:HCl | 34.0 | -24.59 | 39.7 |
| Ethene:HBr | 28.8 | -28.04 | 35.9 |
| [a]Propene:HF | 41.3 | -19.20 | 50.5 |
| [a]Propene:HCl | 35.8 | -20.95 | 41.3 |
| [a]Propene:HBr | 30.0 | -25.97 | 34.5 |
| [m]Propene:HF | 34.2 | -23.68 | 44.0 |
| [m]Propene:HCl | 27.7 | -23.19 | 34.5 |
| [m]Propene:HBr | 21.4 | -29.32 | 28.8 |
| [a]2- Methyl Propene:HF | 43.7 | -16.38 | 52.8 |
| [a]2- Methyl Propene:HCl | 37.1 | -17.74 | 41.7 |
| [a]2- Methyl Propene:HBr | 30.5 | -23.60 | 36.3 |
| [m]2- Methyl Propene:HF | 30.9 | -23.58 | 39.2 |
| [m]2- Methyl Propene:HCl | 22.4 | -21.27 | 28.5 |
| [m]2- Methyl Propene:HBr | 14.1 | -29.96 | 23.9 |

**TABLE III**

Electron density ($\rho(\vec{r}_c)$), laplacian of the electron density ($\nabla^2 \rho(\vec{r}_c)$) and interaction energy ($E_{int}$) for all the π-complexes observed in chlorination of various substituted ethylenes.

| -R | $\rho(\vec{r}_c)$ (e/a$_o^3$) | $\nabla^2 \rho(\vec{r}_c)$ (e/a$_o^5$) | $E_{int}$ (kcal.mol$^{-1}$) |
|---|---|---|---|
| $-NH_2$ | 0.045 | 0.022 | -10.9 |
| $-NHNH_2$ | 0.046 | 0.022 | -10.7 |
| -OH | 0.029 | 0.019 | -6.6 |
| -NHOH | 0.043 | 0.022 | -9.4 |
| $-CH_3$ | 0.022 | 0.016 | -4.8 |
| $-CHCH_2$ | 0.023 | 0.016 | -4.4 |
| -H | 0.019 | 0.015 | -4.0 |
| -CCH | 0.018 | 0.014 | -3.2 |
| -CHO | 0.014 | 0.011 | -2.1 |
| -COOH | 0.014 | 0.011 | -1.9 |
| -COF | 0.012 | 0.009 | -1.4 |
| -CN | 0.012 | 0.010 | -1.4 |

**TABLE IV**

Calculated activation energy ($E^{\ddagger}$), reaction energy ($E_r$), activation free energy ($\Delta G^{\ddagger}$) and free energy of reaction ($\Delta G_r$) values for all the chlorination reactions considered in the study. All the values are in kcal.mol$^{-1}$.

| -R | $E^{\ddagger}$ | $E_r$ | $\Delta G^{\ddagger}$ | $\Delta G_r$ |
|---|---|---|---|---|
| $-NH_2$ | 18.6 | -49.4 | 28.8 | -36.1 |
| $-NHNH_2$ | 18.2 | -50.8 | 28.1 | -37.3 |
| -OH | 28.0 | -49.4 | 37.9 | -35.8 |
| -NHOH | 19.9 | -50.9 | 29.7 | -37.4 |
| $-CH_3$ | 32.1 | -49.5 | 41.0 | -35.9 |
| $-CHCH_2$ | 33.7 | -43.2 | 42.9 | -30.2 |
| -H | 36.2 | -51.2 | 45.8 | -38.3 |
| -CCH | 38.9 | -41.2 | 47.9 | -28.2 |
| -CHO | 43.6 | -42.9 | 52.9 | -29.9 |
| -COOH | 43.9 | -41.6 | 53.3 | -28.6 |
| -NO | 47.0 | -40.1 | 56.1 | -27.6 |
| -COF | 48.4 | -39.4 | 57.8 | -26.6 |
| -CN | 49.1 | -38.7 | 57.9 | -25.8 |
| $-NO_2$ | 51.1 | -40.7 | 60.6 | -27.8 |

**TABLE V**

Calculated electrophilicity values of the reactants ($\omega_R^{(1)}$, $\omega_R^{(2)}$ and $\omega_R^{(3)}$), the TS (ω(TS)) and the product (ω(P)) for all the chlorination reactions considered in the study. All the values are in eV.

| -R | $\omega_R^{(1)}$ | $\omega_R^{(2)}$ | $\omega_R^{(3)}$ | ω(TS) | ω(P) |
|---|---|---|---|---|---|
| $-NH_2$ | 8.876 | 0.600 | 0.394 | 11.754 | 2.213 |
| $-NHNH_2$ | 9.113 | 0.837 | 0.417 | 12.483 | 2.402 |
| -OH | 9.093 | 0.818 | 0.545 | 13.735 | 2.336 |
| -NHOH | 9.562 | 1.287 | 0.532 | 13.481 | 2.680 |
| $-CH_3$ | 9.476 | 1.200 | 0.729 | 11.499 | 2.203 |
| $-CHCH_2$ | 10.359 | 2.084 | 0.588 | 17.543 | 2.630 |
| -H | 9.740 | 1.464 | 0.844 | 7.921 | 2.212 |
| -CCH | 10.527 | 2.252 | 0.671 | 19.709 | 2.808 |
| -CHO | 12.093 | 3.817 | 0.781 | 17.152 | 3.874 |
| -COOH | 11.504 | 3.229 | 0.916 | 14.265 | 3.221 |
| -NO | 14.197 | 5.921 | 0.561 | 25.576 | 6.782 |
| -COF | 12.352 | 4.076 | 1.111 | 17.420 | 3.981 |
| -CN | 11.764 | 3.488 | 0.998 | 21.846 | 3.632 |
| $-NO_2$ | 13.486 | 5.210 | 1.043 | 19.090 | 5.391 |

**TABLE VI**

Calculated activation electrophilicity ($\omega^{\ddagger(1)}, \omega^{\ddagger(2)}, \omega^{\ddagger(3)}$), reaction electrophilicity ($\omega_r^{(1)}, \omega_r^{(2)}, \omega_r^{(3)}$) and activation hardness ($\eta^{\ddagger}$) of all the chlorination reactions considered in the study. [a]All the values are in eV.

| -R | $\omega^{\ddagger(1)}$ | $\omega^{\ddagger(2)}$ | $\omega^{\ddagger(3)}$ | $\omega_r^{(1)}$ | $\omega_r^{(2)}$ | $\omega_r^{(3)}$ | $\eta^{\ddagger}$ |
|---|---|---|---|---|---|---|---|
| $-NH_2$ | 2.879 | 10.761 | 11.361 | -6.662 | 1.613 | 1.819 | -0.158 |
| $-NHNH_2$ | 3.369 | 11.228 | 12.065 | -6.711 | 1.565 | 1.985 | -0.081 |
| -OH | 4.641 | 12.371 | 13.189 | -6.757 | 1.518 | 1.791 | 0.282 |
| -NHOH | 3.919 | 11.662 | 12.949 | -6.882 | 1.393 | 2.148 | 0.173 |
| $-CH_3$ | 2.023 | 9.570 | 10.770 | -7.273 | 1.003 | 1.474 | 0.696 |
| $-CHCH_2$ | 7.184 | 14.871 | 16.955 | -7.729 | 0.546 | 2.042 | 0.529 |
| -H | -1.819 | 5.612 | 7.076 | -7.528 | 0.748 | 1.368 | 0.655 |
| -CCH | 9.182 | 16.787 | 19.039 | -7.719 | 0.556 | 2.137 | 0.711 |
| -CHO | 5.059 | 12.554 | 16.371 | -8.219 | 0.057 | 3.093 | 0.832 |
| -COOH | 2.761 | 10.120 | 13.349 | -8.283 | -0.008 | 2.305 | 1.048 |
| -NO | 11.379 | 19.094 | 25.015 | -7.415 | 0.861 | 6.221 | 0.538 |
| -COF | 5.068 | 12.234 | 16.310 | -8.371 | -0.095 | 2.870 | 1.464 |
| -CN | 10.082 | 17.360 | 20.848 | -8.132 | 0.144 | 2.634 | 1.325 |
| $-NO_2$ | 5.604 | 12.837 | 18.047 | -8.095 | 0.181 | 4.348 | 1.343 |

[a] $\omega_r^{(1,2,3)} = \omega_P - \omega_R^{(1,2,3)}$

$\omega^{\ddagger(1,2,3)} = \omega_{TS} - \omega_R^{(1,2,3)}$

**TABLE VII**

Electrophilicity values (ω) of different species involved in the nitration of toluene and chlorobenzene.

| | Toluene | ω(eV) | Chloro benzene | ω(eV) |
|---|---|---|---|---|
| o-Nitration | o-TS1 + $OH^-$ | 11.928 | o-TS1 + $OH^-$ | 14.688 |
| | o-σ-complex + $OH^-$ | 9.849 | o-σ-complex + $OH^-$ | 13.995 |
| | o-TS2 + $OH^-$ | -0.062 | o-TS2 + $OH^-$ | 0.523 |
| | o-P + $H_2O$ | -23.893 | o-P + $H_2O$ | -22.866 |
| m-Nitration | m-TS1 + $OH^-$ | 14.205 | m-TS1 + $OH^-$ | 21.887 |
| | m-σ-complex + $OH^-$ | 11.457 | m-σ-complex + $OH^-$ | 17.447 |
| | m-TS2 + $OH^-$ | 0.418 | m-TS2 + $OH^-$ | 2.093 |
| | m-P + $H_2O$ | -23.801 | m-P + $H_2O$ | -23.538 |
| p-Nitration | p-TS1 + $OH^-$ | 9.680 | p-TS1 + $OH^-$ | 11.960 |
| | p-σ-complex + $OH^-$ | 8.681 | p-σ-complex + $OH^-$ | 10.523 |
| | p-TS2 + $OH^-$ | 0.112 | p-TS2 + $OH^-$ | 1.772 |
| | p-P + $H_2O$ | -23.872 | p-P + $H_2O$ | -23.628 |

**TABLE VIII**

Calculated reaction electrophilicity ($\omega_r$), reaction energy ($E_r$), activation electrophilicity ($\omega^\ddagger$), activation energy ($E^\ddagger$) and activation hardness ($\eta^\ddagger$) of various nitrotoluenes and nitrochlorobenzenes.

| Species | $\omega_r$ (eV) | $E_r$ (kcal.mol$^{-1}$) | $\omega^\ddagger$ (eV) | $E^\ddagger$ (kcal.mol$^{-1}$) | $\eta^\ddagger$ (eV) |
|---|---|---|---|---|---|
| o-Nitrotoluene | -0.304 | -0.029 | -0.110 | 0.472 | 0.052 |
| m-Nitrotoluene | -0.212 | -0.033 | -1.764 | 1.525 | 0.125 |
| p-Nitrotoluene | -0.282 | -0.033 | 0.016 | 0.244 | 0.018 |
| o-Nitrochlorobenzene | -0.415 | -0.018 | -1.629 | 1.203 | 0.107 |
| m- Nitrochlorobenzene | 0.051 | -0.029 | 9.350 | 8.720 | -0.389 |
| p- Nitrochlorobenzene | -0.039 | -0.030 | -0.574 | 0.698 | 0.072 |

**TABLE IX**

Calculated $\omega_k^-$ values (eV) of toluene and chlorobenzene.

| Toluene | | Chlorobenzene | |
|---|---|---|---|
| 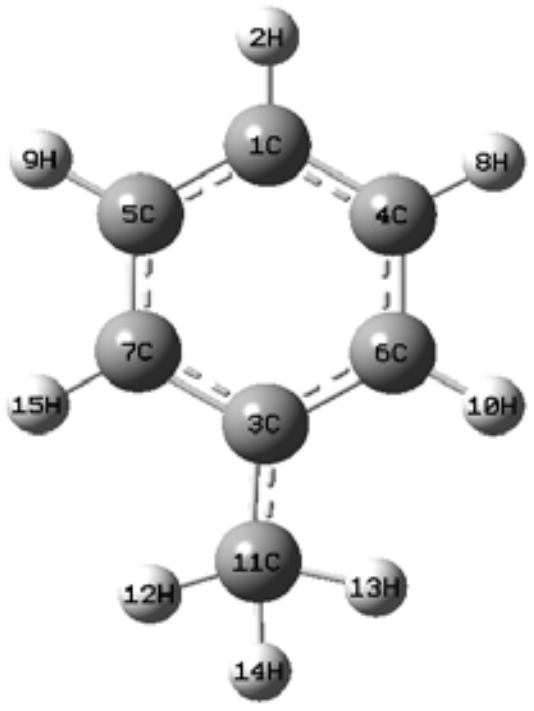 | | 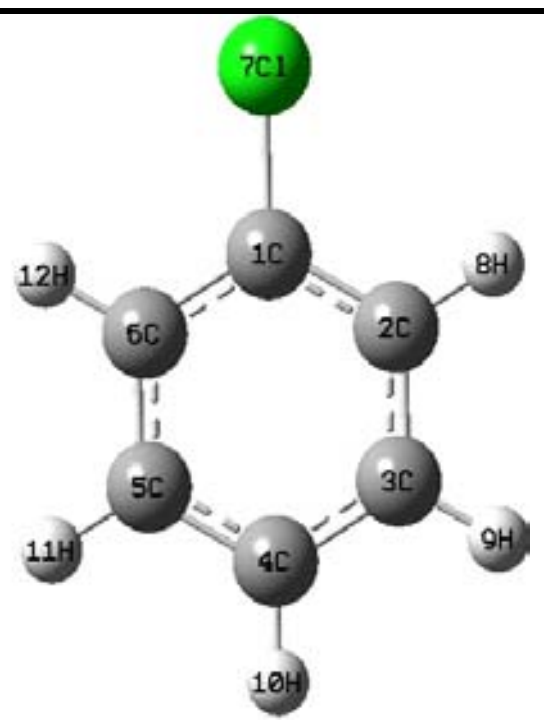 | |
| 1 | 0.461 | 1 | 0.439 |
| 2 | 0.000 | 2 | 0.180 |
| 3 | 0.438 | 3 | 0.103 |
| 4 | 0.108 | 4 | 0.508 |
| 5 | 0.098 | 5 | 0.103 |
| 6 | 0.146 | 6 | 0.180 |
| 7 | 0.156 | 7 | 0.437 |
| 8 | 0.000 | 8 | 0.000 |
| 9 | 0.000 | 9 | 0.000 |
| 10 | 0.000 | 10 | 0.000 |
| 11 | 0.026 | 11 | 0.000 |
| 12 | 0.012 | 12 | 0.000 |
| 13 | 0.008 | | |
| 14 | 0.043 | | |
| 15 | 0.000 | | |

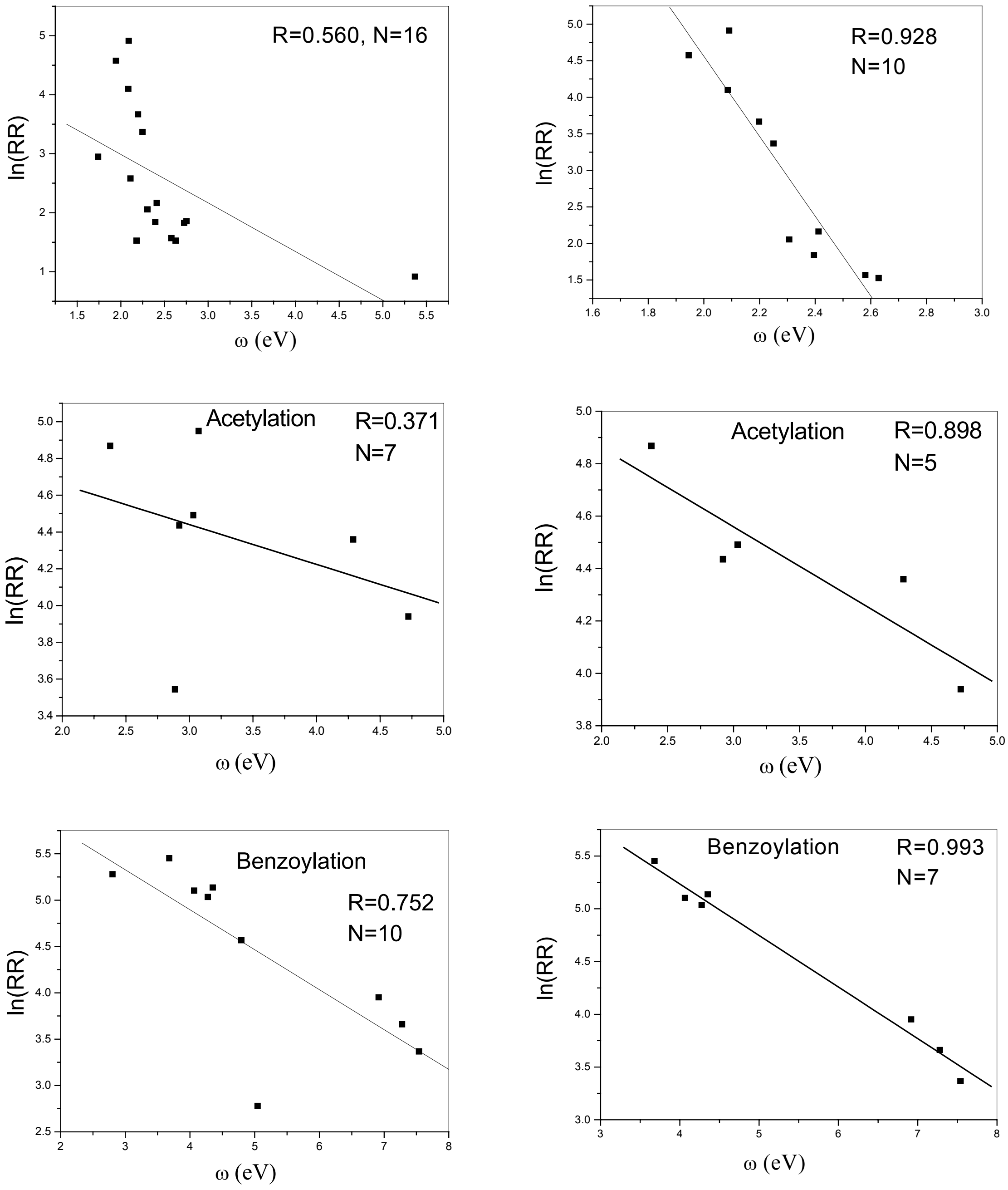

**FIGURE 2.** The experimental ln($RR$) vs. calculated electrophilicity values (eV) for the benzylation, acetylation and benzoylation reactions considered in the study.

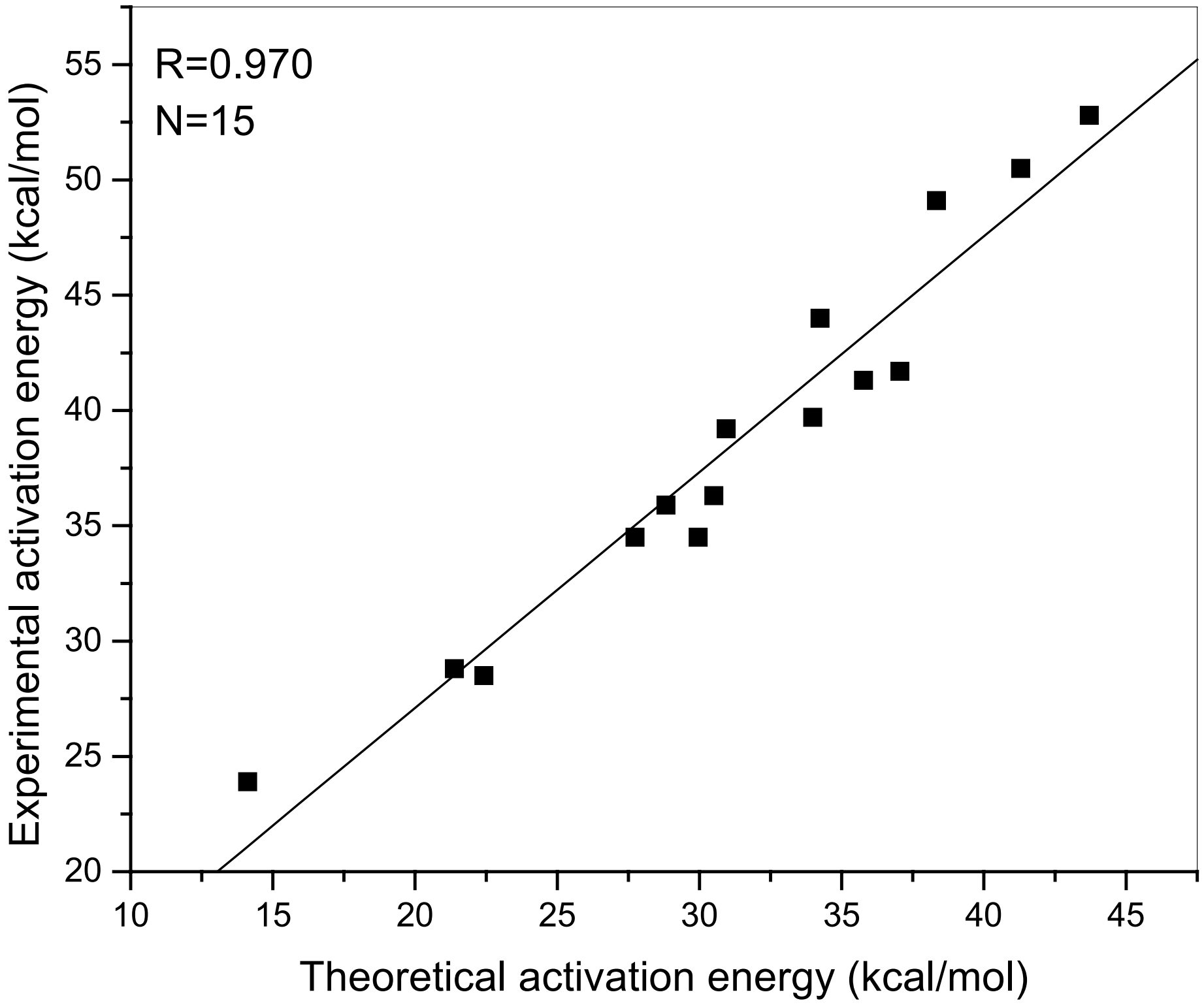

**FIGURE 3.** The theoretical and the experimental activation energies for Markovnikov and anti-Markovnikov addition of hydrogen halides to alkenes considered in the study.

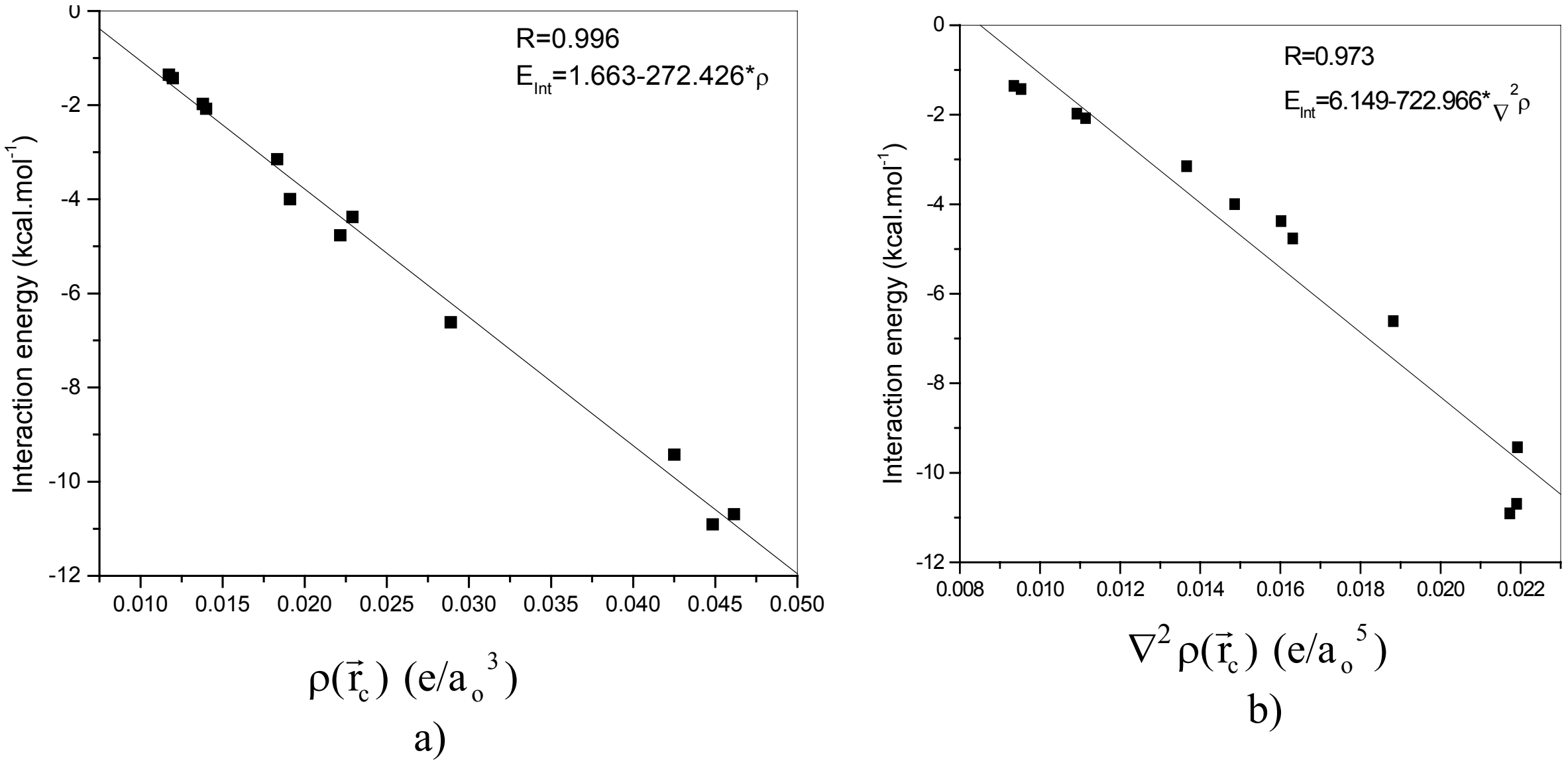

**FIGURE 5**. Plots between a) Interaction energy and $\rho(\vec{r}_c)$ b) Interaction energy and $\nabla^2\rho(\vec{r}_c)$ for all the π-complexes formed between chlorine and π-bond of the substituted ethylenes.

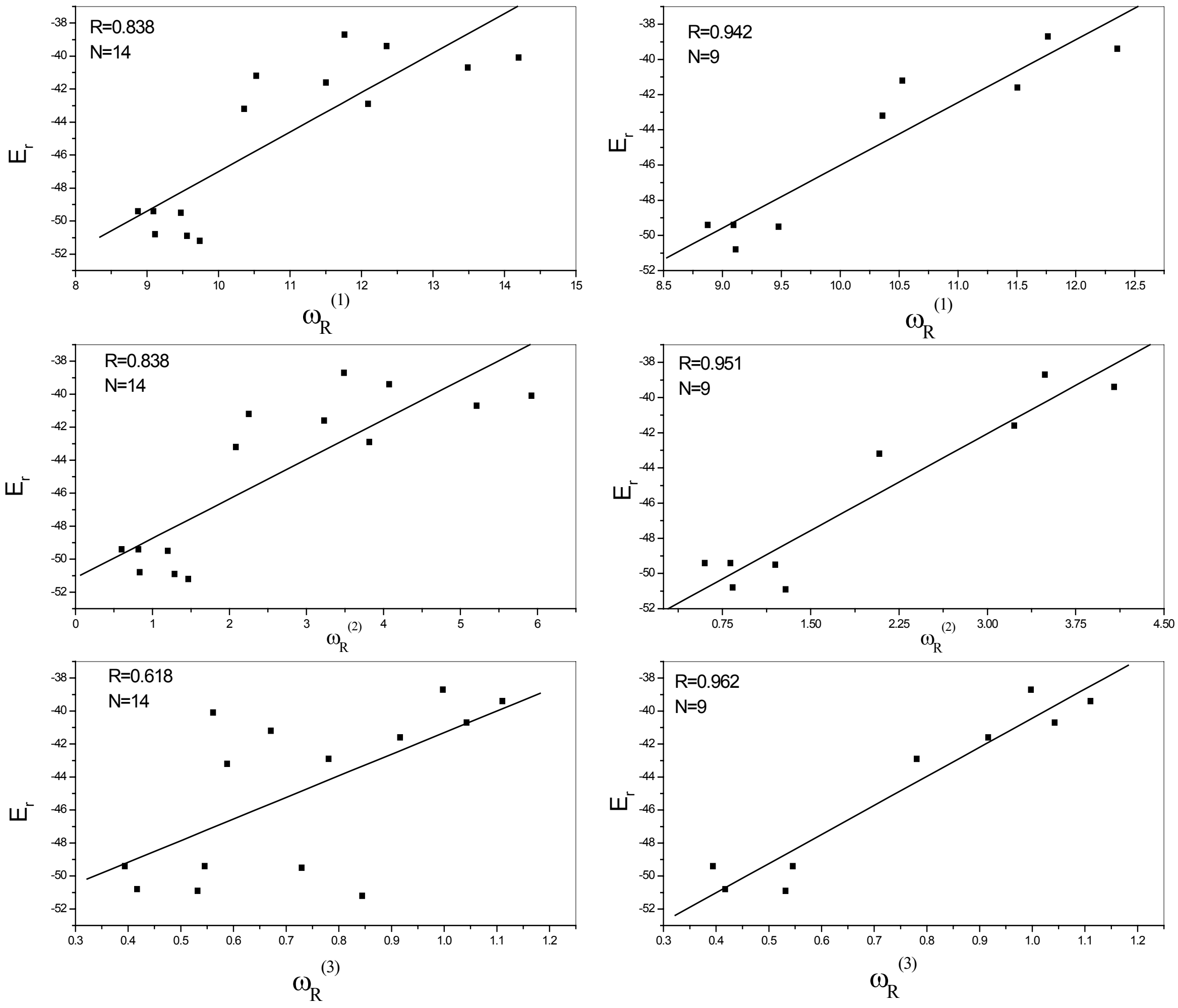

**FIGURE 6.** Plots of $E_r$ vs $\omega_R^{(1,2,3)}$

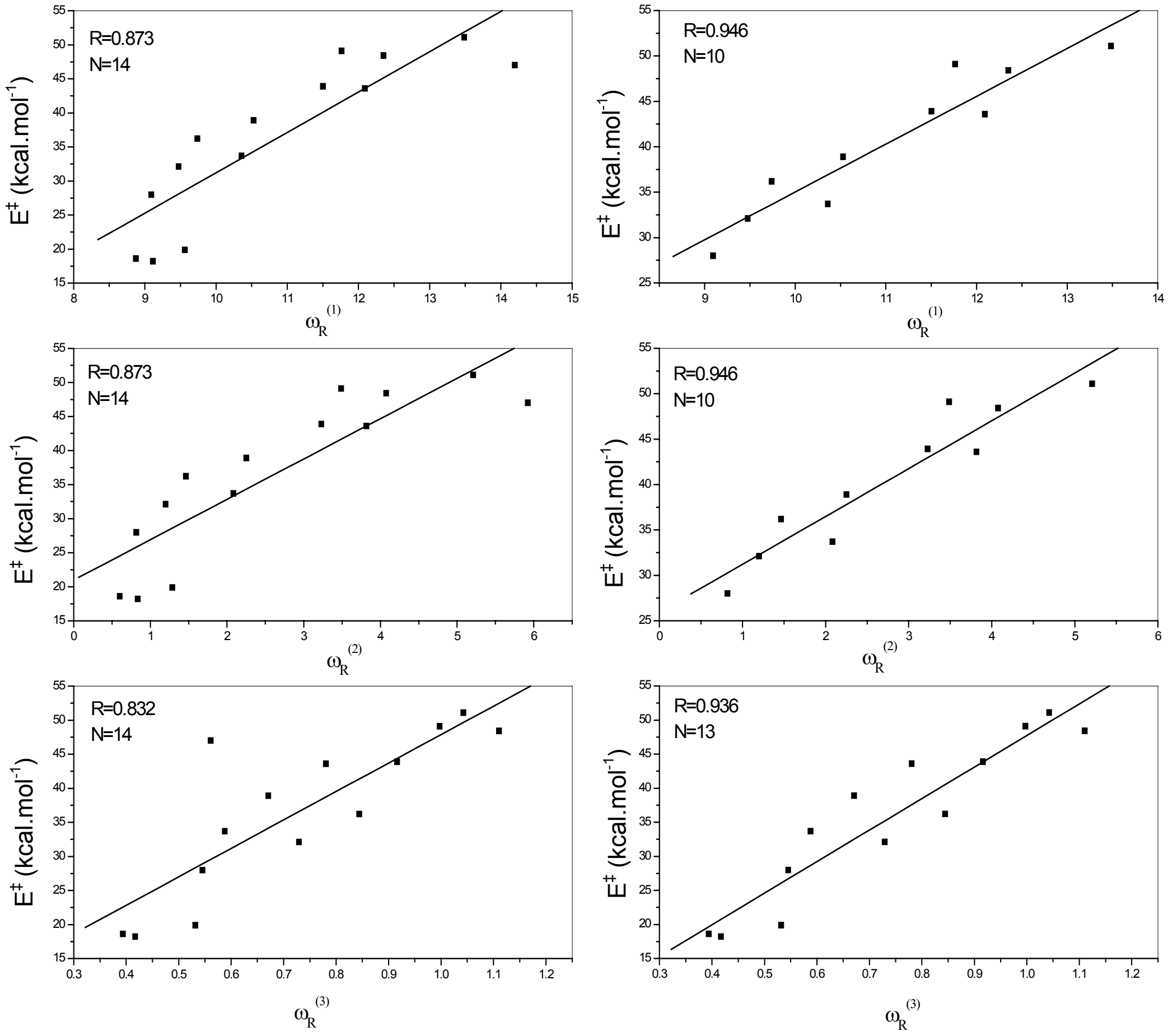

**FIGURE 6.** Plots of $E^{\ddagger}$ vs $\omega_R^{(1,2,3)}$

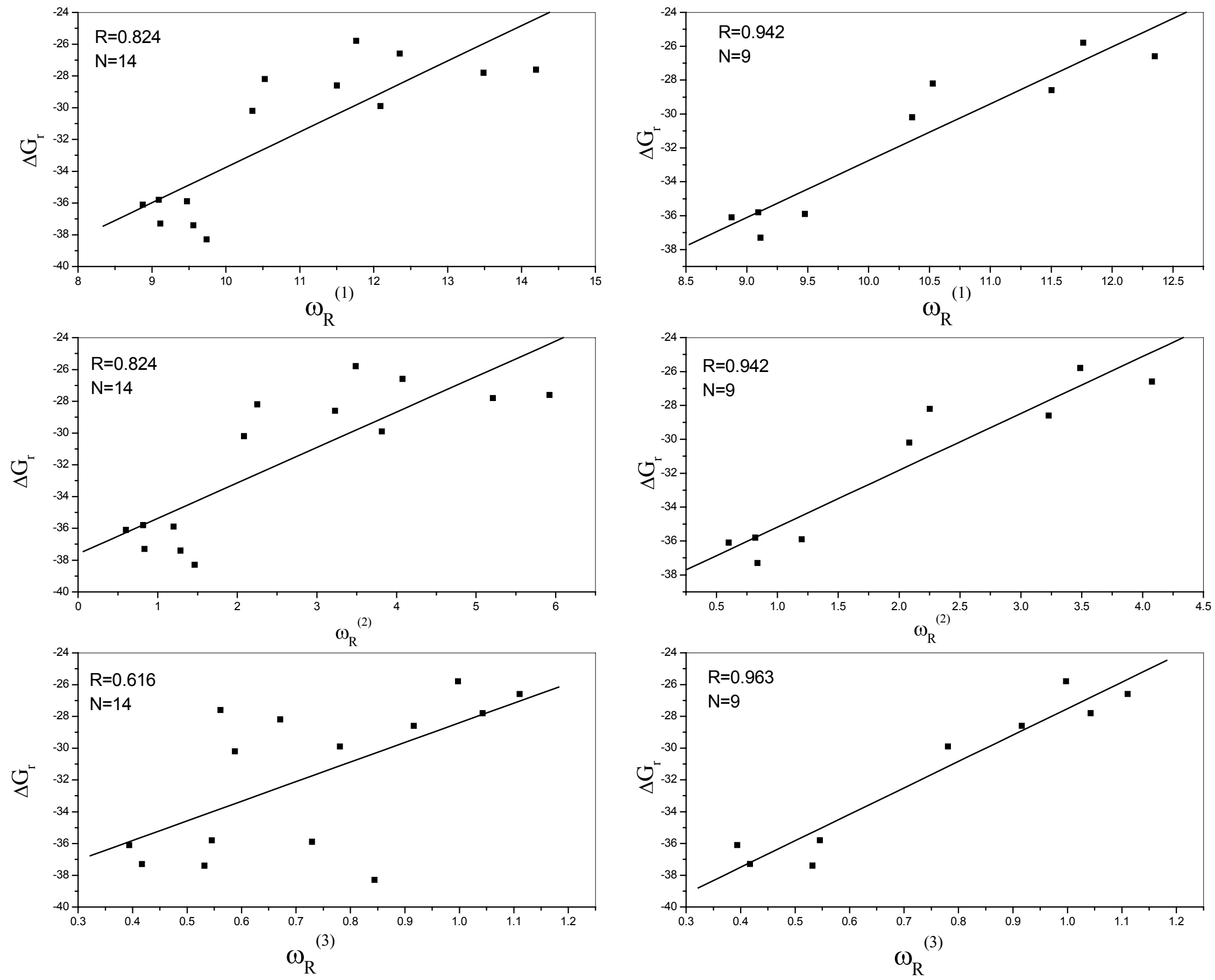

**FIGURE 6.** Plots of $\Delta G_r$ vs $\omega_R^{(1,2,3)}$

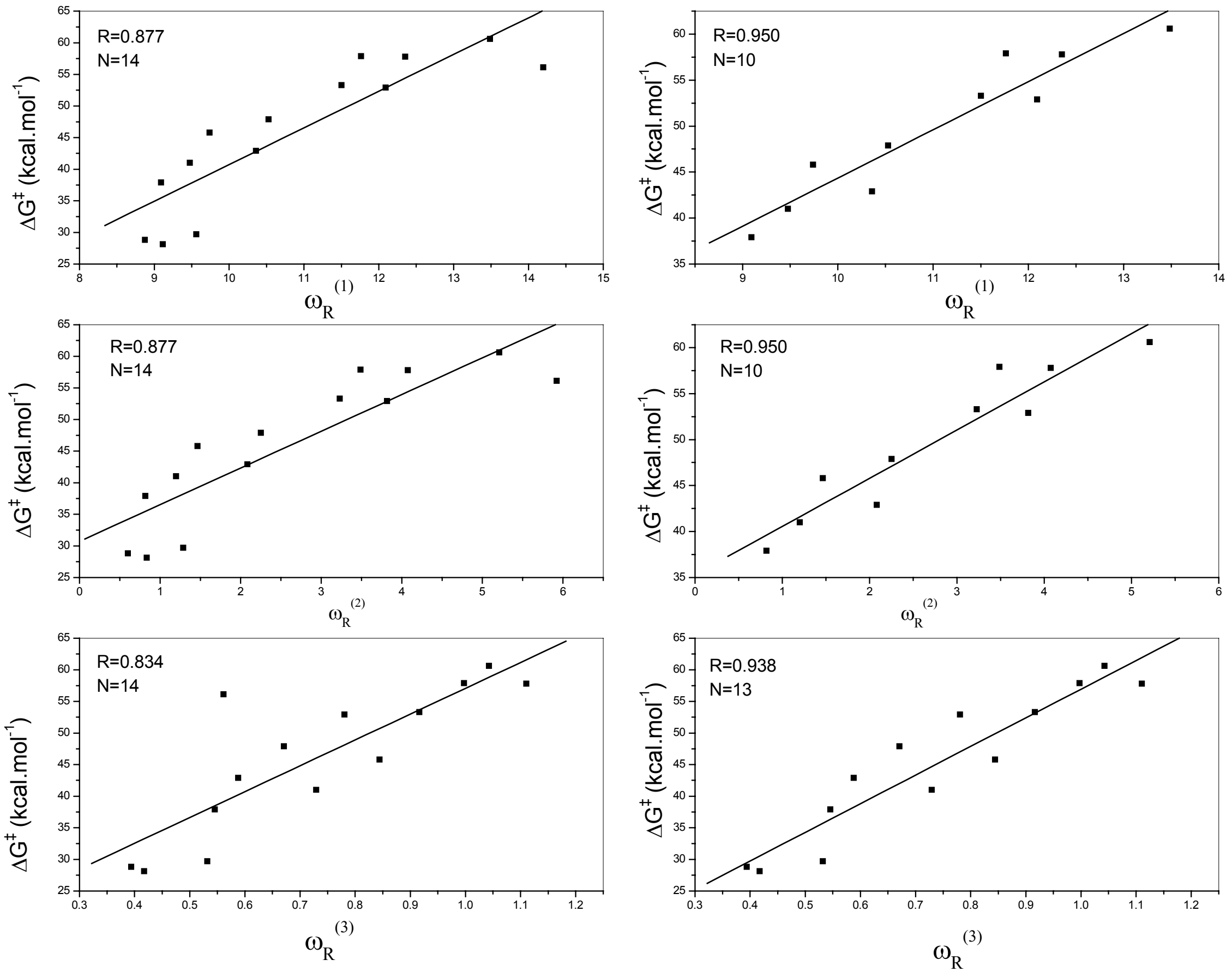

**FIGURE 6.** Plots of $\Delta G^{\ddagger}$ vs $\omega_R^{(1,2,3)}$

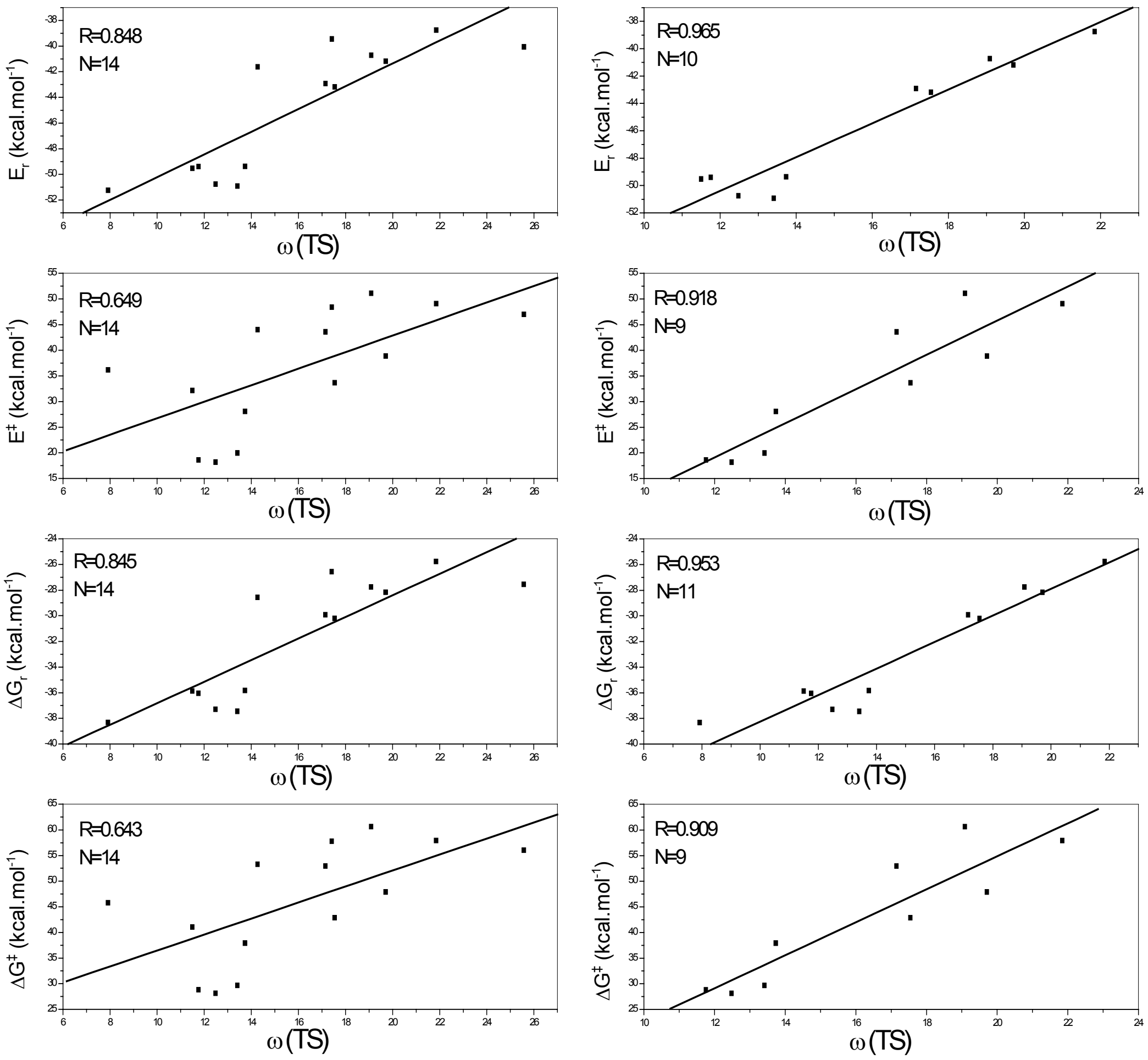

**FIGURE 6.** Plots of Y (Y= $E_r$, $E^{\ddagger}$, $\Delta G_r$, $\Delta G^{\ddagger}$) vs $\omega(TS)$.

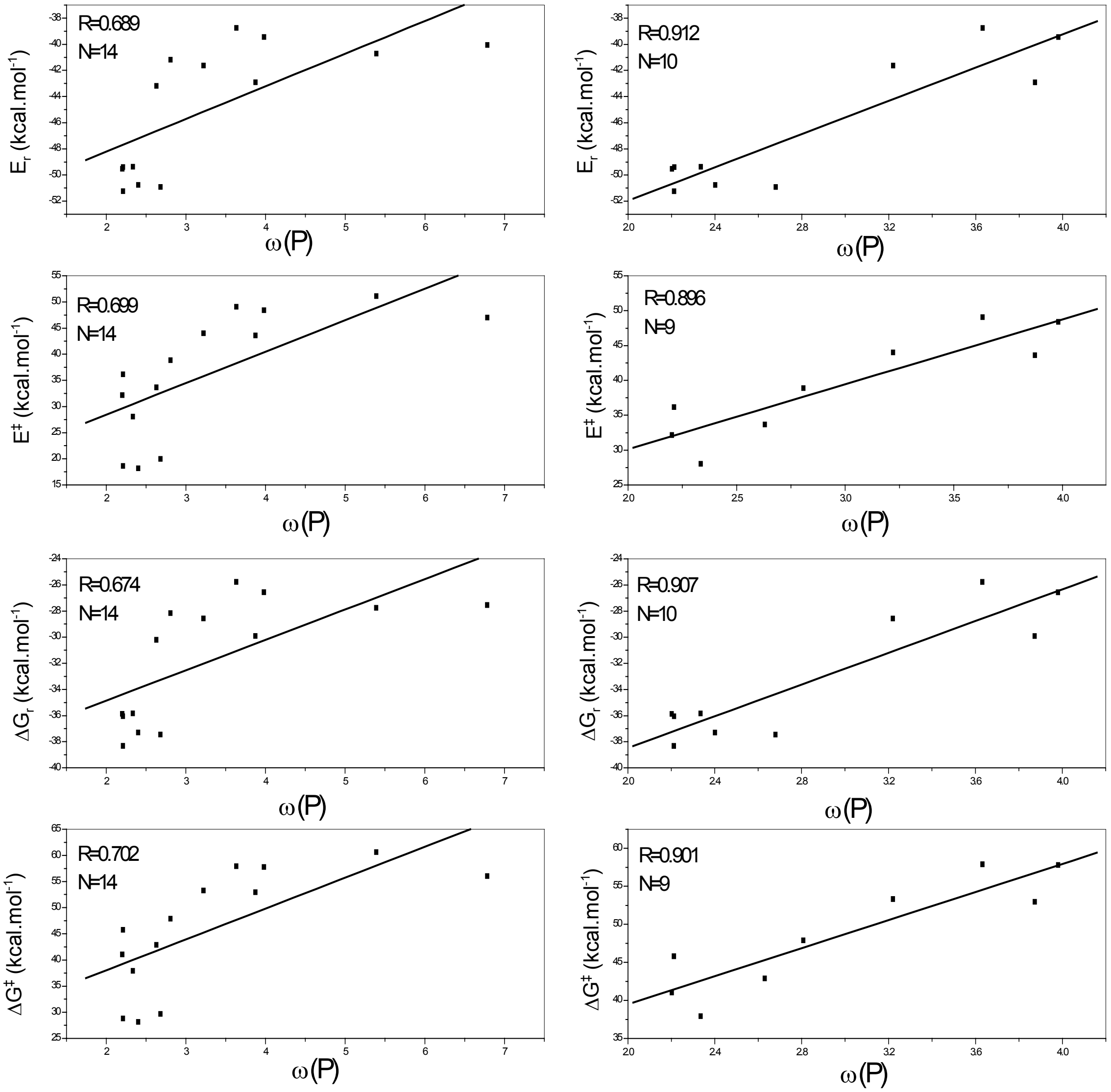

Plots of Y (Y= $E_r$, $E^{\ddagger}$, $\Delta G_r$, $\Delta G^{\ddagger}$) vs $\omega(P)$

**FIGURE 6.** The correlations of various kinetic and thermodynamic quantities with $\omega_R^{(1,2,3)}$, $\omega_{TS}$ and $\omega_P$ (eV) for the chlorination reactions considered in the present study.

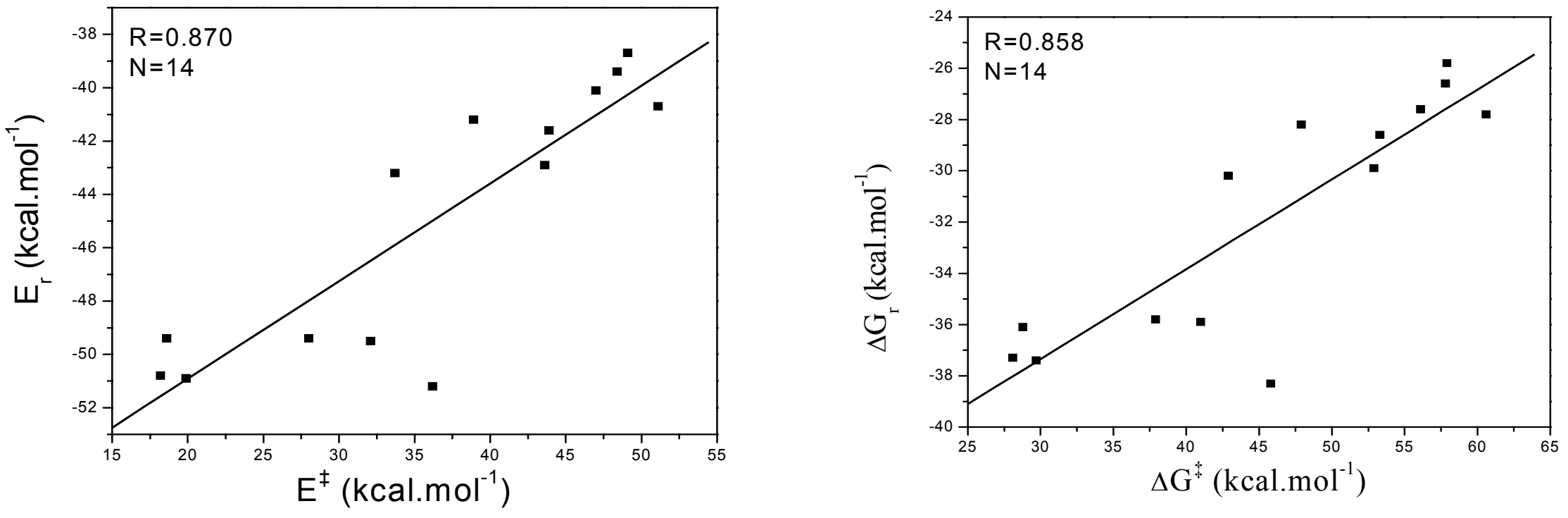

a B

**FIGURE 7.** Plot of a) reaction energy versus activation energy and b) reaction free energy versus free energy of activation for the chlorination reactions considered in the present study.

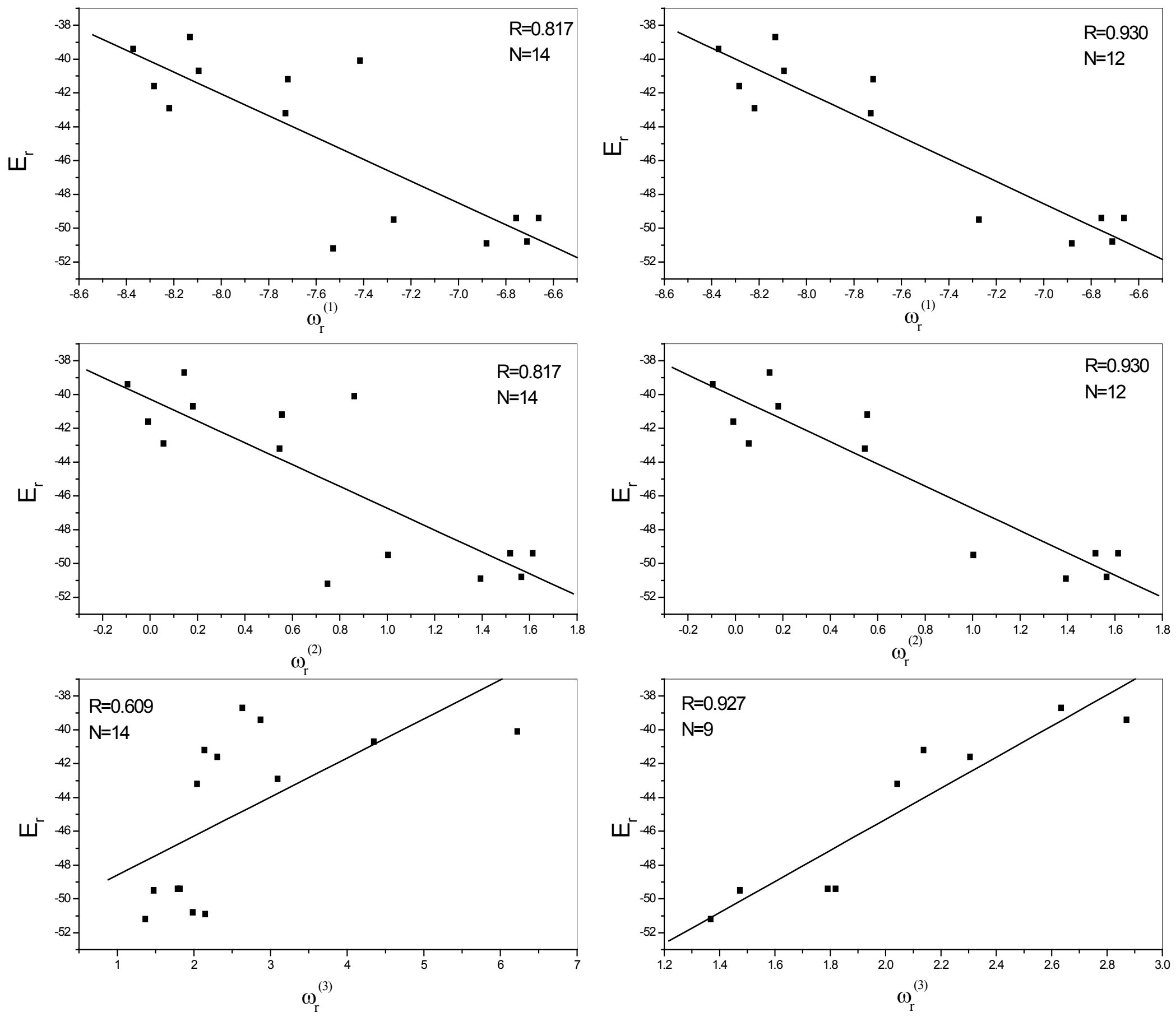

**FIGURE 8a.** Plots of $E_r$ vs $\omega_r^{(1,2,3)}$ .

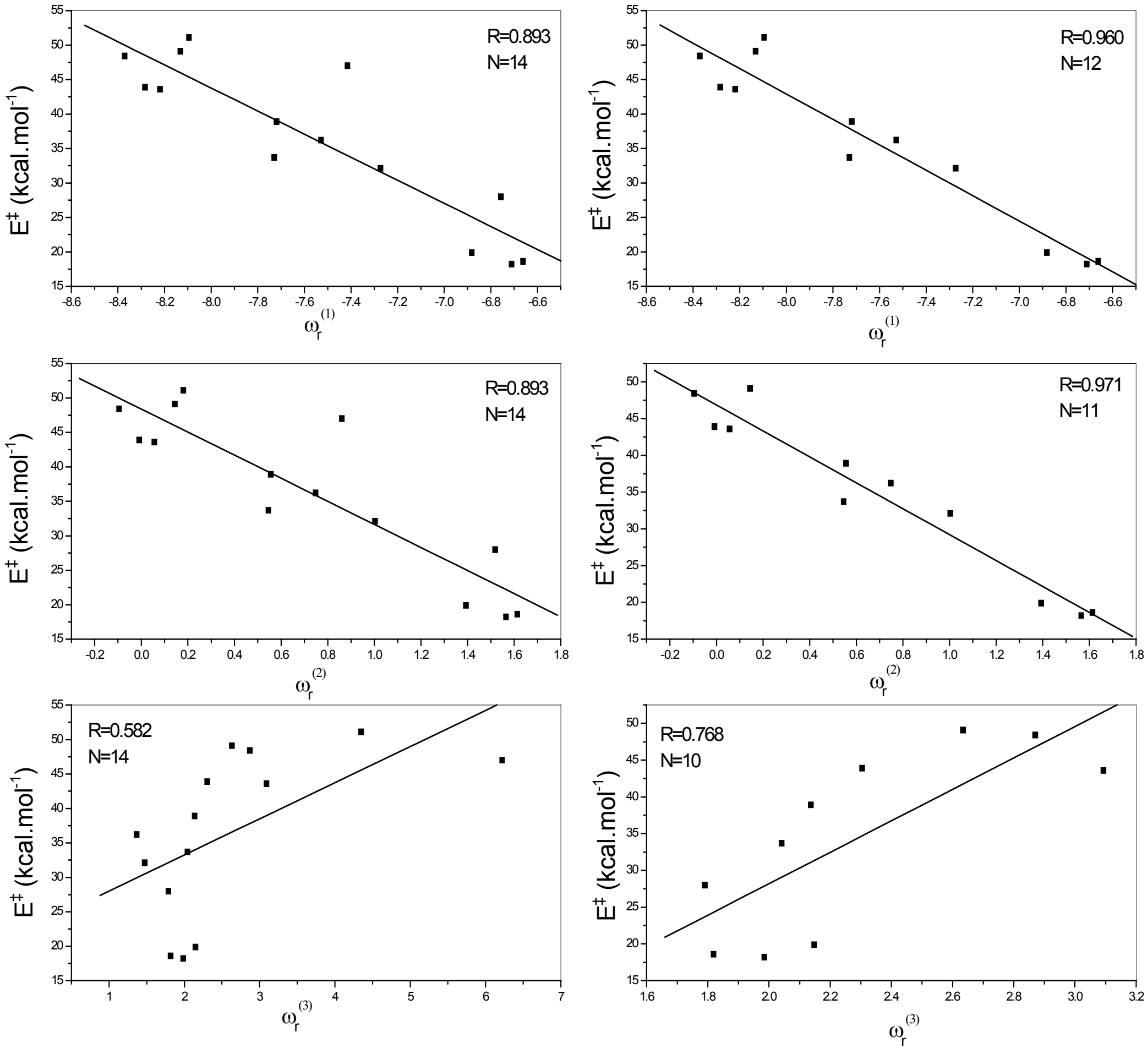

**FIGURE 8b.** Plots of $E^{\ddagger}$ vs $\omega_r^{(1,2,3)}$ .

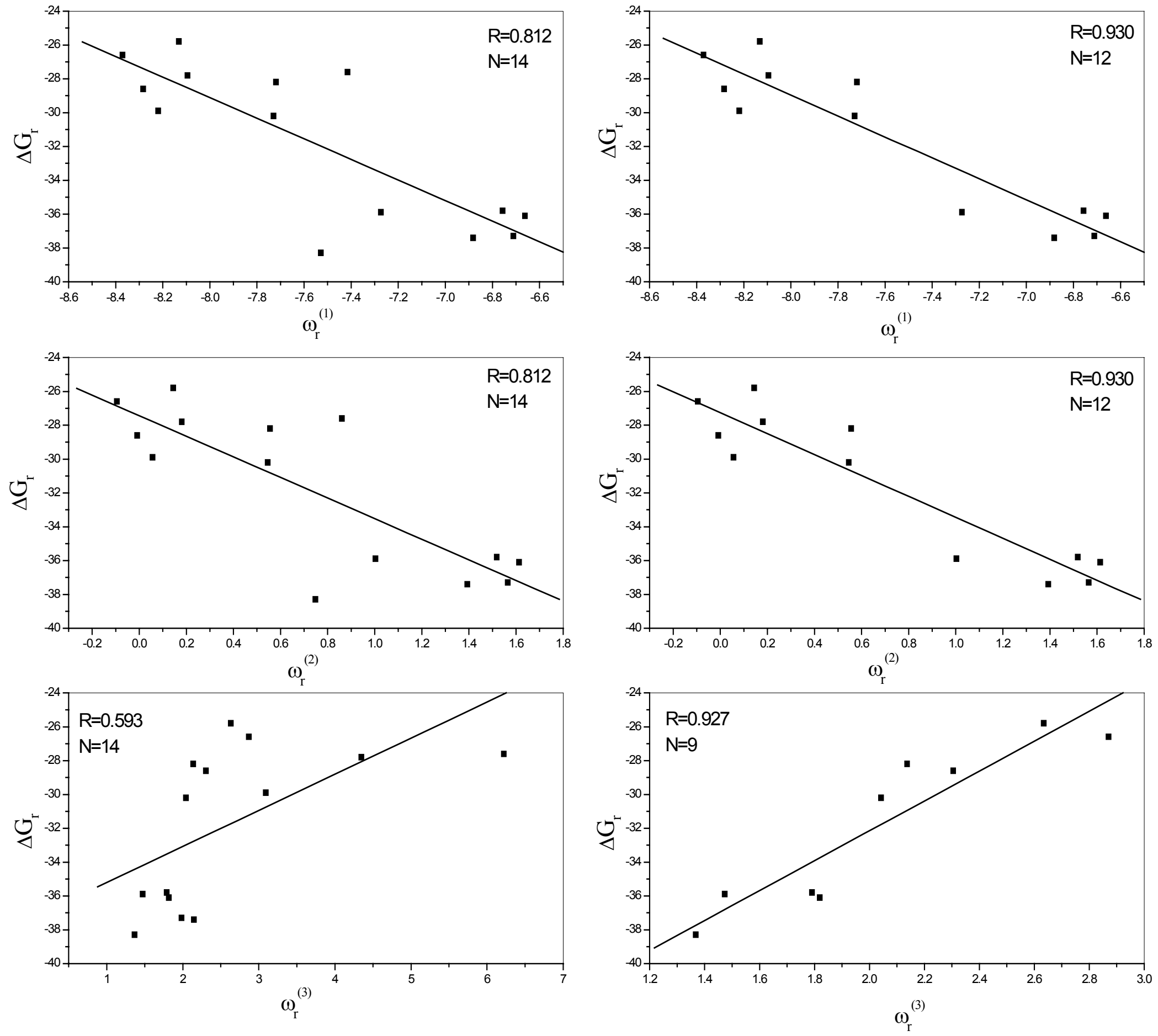

**FIGURE 8c.** Plots of $\Delta G_r$ vs $\omega_r^{(1,2,3)}$.

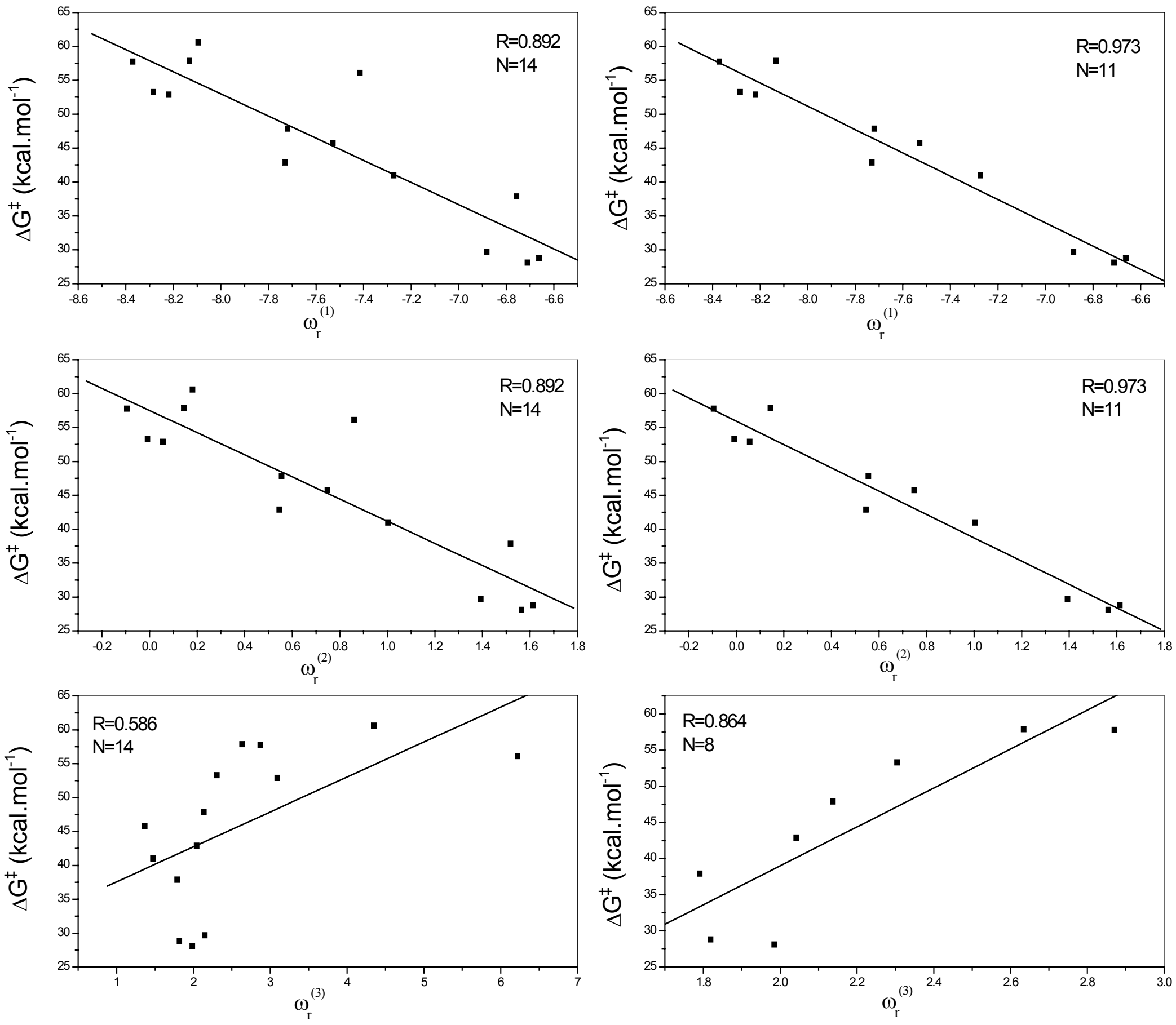

**FIGURE 8d.** Plots of $\Delta G^{\ddagger}$ vs $\omega_r^{(1,2,3)}$ .

**FIGURE 8.** Plots of a) reaction energy, b) activation energy, c) reaction free energy and d) free energy of activation with reaction electrophilicity $\omega_r^{(1,2,3)}$ for the chlorination reactions.

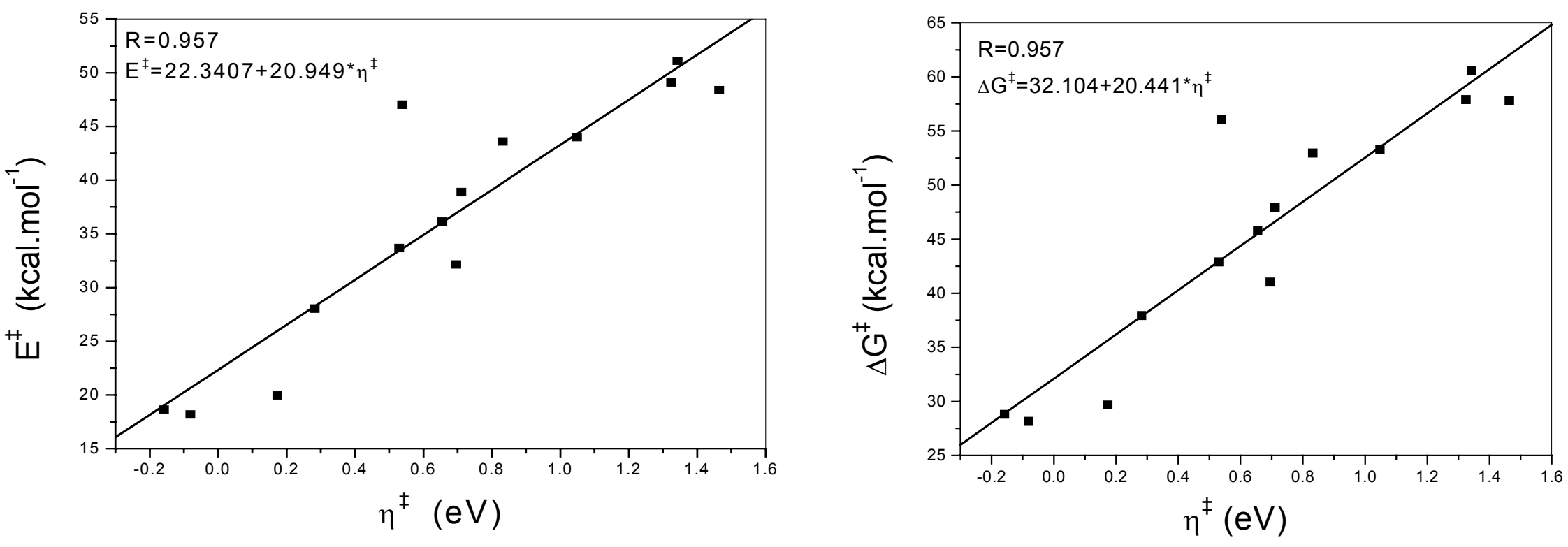

(a) (b)

**FIGURE 9**. Plot of a) activation energy with activation hardness, b) free energy of activation with activation hardness for the chlorination reactions.

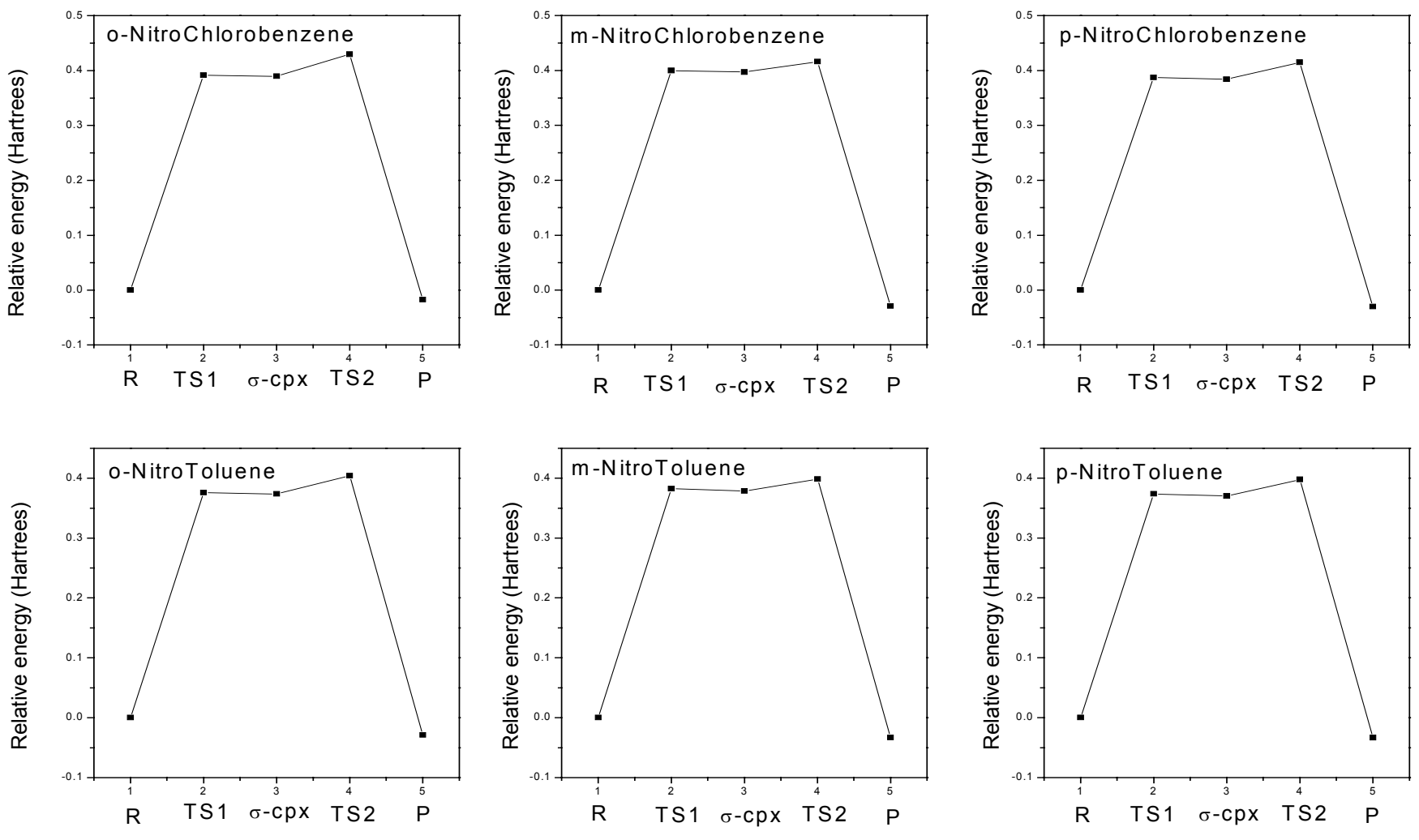

**FIGURE 11.** Relative energies of various reactants, transition states, products and $\sigma$-complexes associated with the nitration of toluene and chlorobenzene. Here cpx = complex.

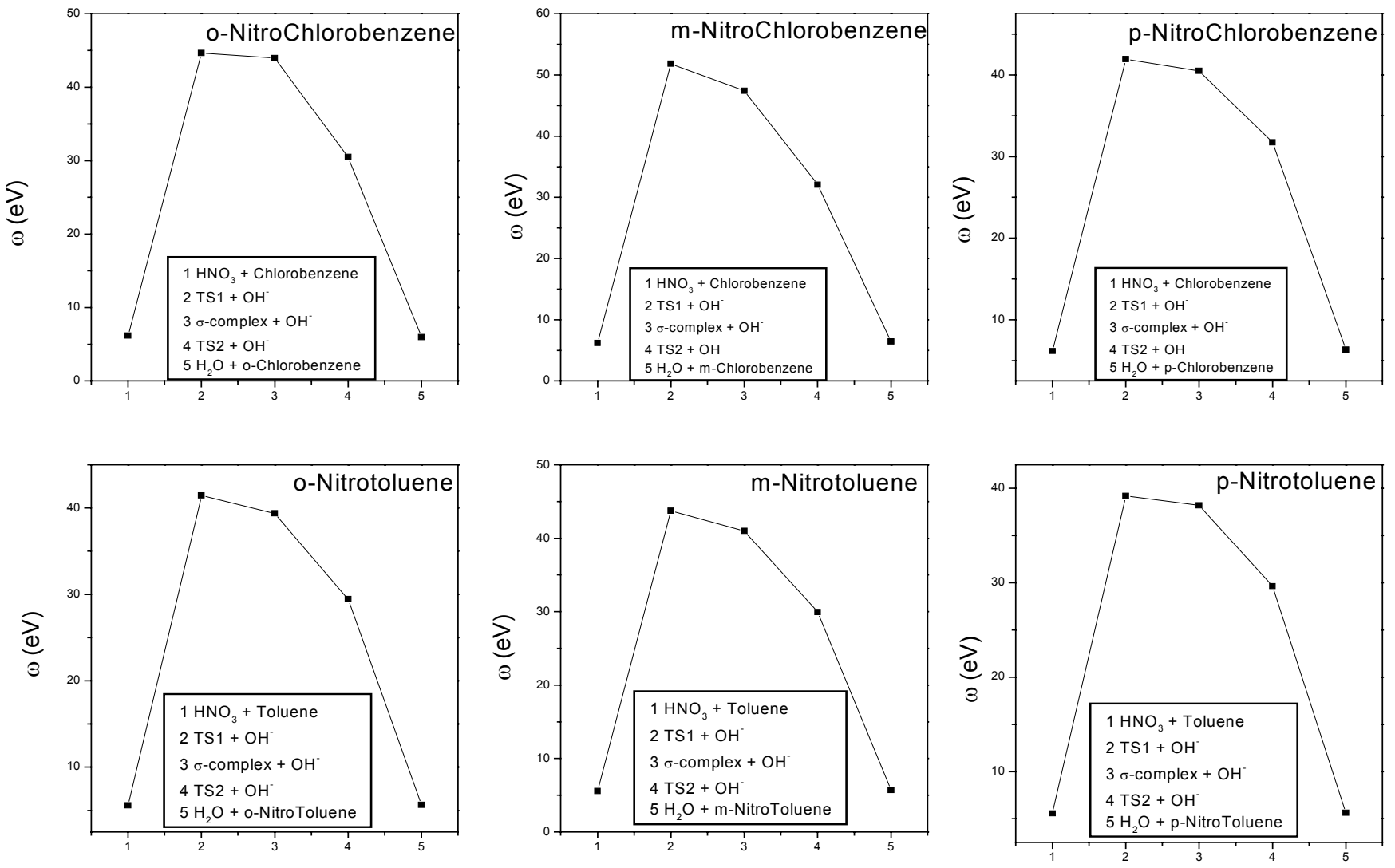

**FIGURE 12.** Electrophilicities of various reactants, transition states, products and $\sigma$ - complexes associated with the nitration of toluene and chlorobenzene.